\documentclass[12pt]{article}
\usepackage{amsmath}
\usepackage{graphicx}
\usepackage{natbib}
\usepackage{url} 

\newcommand{\blind}{0}

\usepackage{algorithm,hyperref,natbib,microtype,xr-hyper}
\newcommand{\add}[1]{{#1}}  

\date{}

\begin{document}

\def\spacingset#1{\renewcommand{\baselinestretch}%
{#1}\small\normalsize} \spacingset{1}


\if0\blind
{
  \title{\bf The $G$-Wishart Weighted Proposal Algorithm:
Efficient Posterior Computation for Gaussian Graphical Models}
  \author{Willem van den Boom\thanks{
    Email: \href{mailto:vandenboom@nus.edu.sg}{vandenboom@nus.edu.sg}. This work was supported by the Singapore Ministry of Education Academic Research Fund Tier~2 under Grant MOE2019-T2-2-100.}\hspace{.2cm}\\
    \add{Yong Loo Lin School of Medicine}, National University of Singapore\\
    and \\
    Alexandros Beskos \\
    Department of Statistical Science, University College London\\
    Alan Turing Institute, UK\\
    and \\
    Maria De Iorio \\
    Yong Loo Lin School of Medicine,
    National University of Singapore \\
    Singapore Institute for Clinical Sciences, A*STAR\\ 
    Department of Statistical Science, University College London
    }
  \maketitle
} \fi

\if1\blind
{
  \bigskip
  \bigskip
  \bigskip
  \begin{center}
    {\LARGE\bf The $G$-Wishart weighted proposal  algorithm:
Efficient posterior computation for Gaussian graphical models}
\end{center}
  \medskip
} \fi

\bigskip
\begin{abstract}  
Gaussian graphical models can capture complex dependency structures among variables. For such models, Bayesian inference is attractive as it provides principled ways to incorporate prior information and to quantify uncertainty through the posterior distribution. However, posterior computation under the conjugate $G$-Wishart prior distribution on the precision matrix is expensive for general non-decomposable graphs. We therefore propose a new Markov chain Monte Carlo (MCMC) method named the $G$-Wishart weighted proposal algorithm (WWA). WWA's distinctive features include delayed acceptance MCMC, Gibbs updates for the precision matrix and an informed proposal distribution on the graph space that enables embarrassingly parallel computations. Compared to existing approaches, WWA reduces the frequency of the relatively expensive sampling from the $G$-Wishart distribution. This results in faster MCMC convergence, improved MCMC mixing and reduced comput\add{ing} time. Numerical studies on simulated and real data show that WWA provides a more efficient tool for posterior inference than competing state-of-the-art MCMC algorithms.
\add{Supplemental materials for the article are available online.}
\end{abstract}

\noindent%
{\it Keywords:}  
Exchange algorithm;
Hyper \add{I}nverse Wishart distribution;
Locally balanced proposal;
Reversible jump MCMC;
Scalable Bayesian computations
\vfill

\newpage
\spacingset{1.5} 

\newpage
\spacingset{1.5} 

\section{Introduction}
\label{sec:intro}

Gaussian graphical models \citep[GGMs,][]{Dempster1972,Lauritzen1996}
are a powerful tool to investigate conditional independence structure among variables \add{which are} represented by the nodes of \add{a} graph. Graph estimation is often very challenging given the dimensionality of the graph space. In the frequentist literature, different strategies have been proposed to bypass this problem. For example,
\citet{Friedman2007} propose the graphical lasso, which involves the estimation of the precision matrix of a \add{M}ultivariate Gaussian vector through penalised likelihood methods. Then, from the estimation of the covariance, the graph is constructed by drawing an edge between variables whose partial correlation is estimated as different from zero. Alternatively,
nodewise regression \citep{Zhou2011} approximates the joint distribution of the variables by considering individual regressions of each variable on the others. This leads to a computationally efficient set\add{-}up which is also amenable to parallelisation, at the cost of not being founded upon a probabilistically consistent model.

On the other hand, in the Bayesian framework, graph estimation requires the specification of a prior on the space of graphs and, conditionally on the graph, a prior on the precision matrix.
\add{
 Bayesian inference allows for principled uncertainty quantification, 
handling complexity through the specification of (conditionally independent) submodules and superior performance compared to frequentist methods.
The posterior distribution on the graph space provided by the Bayesian framework provides principled and interpretable measures of uncertainty such as posterior edge inclusion probabilities.
This contrasts with many frequentist methods such as the graphical lasso which conceptually focus on the precision matrix.
Then, graph estimation happens through thresholding or penalisation of this matrix,
which do not provide interpretable measures of uncertainty on the graph space.
Additionally, the modular nature of MCMC enables
its use in extended models that incorporate GGMs
such as sparse seemingly unrelated regressions \citep[e.g.,][]{Wang2010b,Bhadra2013}
while propagating the uncertainty of the GGM to other parts of the model.
Finally,
the empirical comparisons in \citet{Mohammadi2015}
show that Bayesian inference with the $G$-Wishart prior considered in this work
often outperforms frequentist methods such as the graphical lasso
in terms of graph structure recovery and precision matrix estimation. 

The previous points provide the rationale for this work, as the goal is to better enable and speed up Bayesian inference in GGMs.}
Posterior inference is \add{usually} performed through Markov chain Monte Carlo \citep[MCMC, e.g.,][]{Wang2012,Hinne2014} and more recently sequential Monte Carlo \citep[SMC,][]{Tan2017,vandenBoom2022}. However, these methods are often associated with high computational cost.
A possible solution is to constrain the analysis to decomposable graphs \citep[e.g.,][]{Giudici1999,Letac2007,Scott2008,Wang2010b,Bornn2011,Bhadra2013}
or to related subsets of the graph space \citep{Khare2018}
as the associated distributions are tractable.
The assumption of decomposability is hard to justify from an applied perspective
and increasingly restrictive as the number of nodes increases.
In the large data limit where the posterior concentrates, decomposability results in
spurious edges which constitute a
minimal triangulation of the true graph \citep{Fitch2014,Niu2021}. In practice, this implies that up to half of the edges in the estimated graph can be spurious, even if the posterior is highly concentrated.

To avoid the assumption of decomposable graphs, \citet{Roverato2002} introduces the $G$-Wishart prior distribution for the precision matrix conditional on a graph, which is an extension of the Hyper Inverse Wishart distribution \citep{Dawid1993} employed in the case of decomposable graphs. This allows for more flexibility at the cost of more expensive computations. 
Given the difficulties \add{in exploring} the posterior space,
stochastic search methods
in a Bayesian model have been developed
\citep{Jones2005,Scott2008,Lenkoski2011}, which aim to identify graphs with high posterior probability.
Nevertheless, 
MCMC algorithms have received most attention in the literature \citep[e.g.,][]{Wang2012,Cheng2012,Lenkoski2013,Mohammadi2015} as they allow for full posterior inference.

Inference with the $G$-Wishart distribution is challenging. For instance, \citet{Wang2015}, \citet{Gan2018}, \citet{Li2019} and \citet{Sagar2021} obtain major improvements in computational efficiency by replacing the $G$-Wishart distribution with shrinkage priors on the precision matrix that enable fast Gibbs sampling updates or EM algorithms.
Moreover, building on MCMC methods,
SMC \citep{Tan2017} and unbiased Monte Carlo approximation \citep{vandenBoom2022}
have also been considered as these techniques are \textit{embarrassingly parallel}.

Still working with a $G$-Wishart prior, we propose an MCMC algorithm by carefully addressing the major computational bottlenecks in the current literature.
Our work builds on advances in MCMC algorithms, in particular, work on delayed Metropolis-Hastings acceptance \citep{Christen2005} and informed proposals \citep{Zanella2019} 
which we extend to the graph literature. 
First, we propose a delayed acceptance MCMC step to reduce the number of times we need to sample from the $G$-Wishart distribution. Such sampling involves iterating a\add{n} $O(pd^3)$ algorithm to convergence where $p$ is the number of nodes and $d$ the degree of the graph \citep{Lenkoski2013}. 
Moreover, we introduce Gibbs updates for the precision matrix enabled by a node reordering which further reduce the need to sample from the $G$-Wishart distribution. Finally,
we develop an informed proposal distribution for graphs
which enables the use of parallel computing environments still in an MCMC framework. As the main distinctive features of the proposed method relate to its proposal distribution on the graph space, we refer to it as the $G$-Wishart weighted proposal algorithm (WWA). We show that WWA improves computation significantly and allows for exploration of larger graph spaces.

The paper is structured as follows. Section~\ref{sec:model} introduces Bayesian GGMs based on the $G$-Wishart prior, while Section~\ref{sec:post} reviews related literature on posterior inference.
Section~\ref{sec:wwa} describes the proposed WWA and 
contextualises it.
Section~\ref{sec:empirical} presents simulation studies to investigate the performance of WWA and compares it with the state of the art. In Section~\ref{sec:gene}, we consider a real data application.
We conclude the paper in 
Section~\ref{sec:discussion}.

\subsection{Model Description}
\label{sec:model}

\add{Object} of inference is a graph $G=(V,E)$ defined by a set of edges $E \subset {\{(i,j)\mid 1\leq i<j\leq p \}}$ 
that represent links among the nodes in $V=\{1,\ldots,p\}$.
In the GGM framework,
we have an
$n\times p$ data matrix $Y$ with independent rows $Y_i$, $i=1,\ldots,n$, corresponding to a $p$-dimensional random vector with its elements represented by nodes on the graph. Each $Y_i$ is distributed according to a Multivariate Gaussian distribution $\mathcal{N}(0_{p\times 1},\, K^{-1})$
with precision matrix $K$. We assume that the precision matrix $K$ depends on the graph $G$:
we have that $K_{ij} = 0$
if nodes $i$ and $j$ are not connected, while if there is an edge between two nodes in the graph then the corresponding element of the precision matrix is different from zero with probability one.
Thus,
$K\in M^+(G)$ where $M^+(G)$ is the cone of positive-definite matrices $K$ with $K_{ij} = 0$ for $(i,j)\notin E$.
\add{The graph} $G$ determines the conditional independence structure of the $p$ variables in $Y_i$, $i=1,\ldots,n$,
since
$K_{ij} = 0$
implies that the $i$-th and $j$-th columns of $Y$ are independent
conditionally on the others
by properties of the Multivariate Gaussian distribution.

A popular choice as prior for the precision matrix $K$ conditional on the graph $G$ is the $G$-Wishart distribution $\mathcal{W}_G(\delta, D)$ as it induces conjugacy and allows one to work with non-decomposable graphs \citep{Roverato2002}. It is paramet\add{e}rised by degrees of freedom $\delta > 2$
and a positive-definite rate matrix $D$. Its density is
\[
	p(K\mid G) = \frac{1}{I_G(\delta, D)} |K|^{\delta/2 - 1} \exp\left\{ -\tfrac{1}{2} \mathrm{tr}(K^\top D) \right\},
	\quad K\in M^+(G),
\]
where
$I_G(\delta, D)$ is a normalising constant.
Due to conjugacy, ${K\mid G, Y} \sim {\mathcal{W}_G(\delta^\star,\, D^\star)}$
where $\delta^\star = \delta+n$, $D^\star = D+Y^\top Y$.
The model is completed by specifying a prior $p(G)$ on the graph space. We highlight that the following development does not assume any particular form for $p(G)$.

\subsection{Posterior Distribution}
\label{sec:post}

The goal is to compute the posterior distribution
\citep[e.g.,][]{AtayKayis2005}
\[
	p(G\mid Y)\propto p(G) \int_{M^+(G)} p(K\mid G)\, p(Y\mid K)\, dK
	= \frac{p(G)\, I_G(\delta^\star, D^\star)}{(2\pi)^{np/2} I_G(\delta, D)}.
\]
The normalising constant
$I_G(\delta, D)$ does not have a simple analytical form for general non-decomposable $G$, making evaluation of a Metropolis-Hastings acceptance probability infeasible.
To overcome this problem,
Monte Carlo \citep{AtayKayis2005}
and Laplace \citep{Moghaddam2009,Lenkoski2011} approximations
of $I_G(\delta, D)$
have been developed. Alternatively,
\citet{Uhler2018} provide a recursive expression for $I_G(\delta, D)$, but it results in a computationally efficient procedure only for specific types of graphs.

Another line of work avoids direct evaluation of $I_G(\delta, D)$
through application of the exchange algorithm \citep{Murray2006} within a broader MCMC. \citet{Wang2012} and \citet{Cheng2012} employ this strategy in which sampling from the $G$-Wishart is performed through \add{the edgewise or the maximum clique block Gibbs sampler described in \citet{Wang2012}}.
More recently, \citet{Hinne2014}, \citet{Mohammadi2015} and \citet{vandenBoom2022} propose methodology based on 
the exchange algorithm where sampling from
$\mathcal{W}_G(\delta, D)$ is performed through the exact $G$-Wishart sampler by \citet{Lenkoski2013}, making the algorithm more accurate in terms of exploration of posterior space.

\section{WWA: \texorpdfstring{$G$}{G}-Wishart Weighted Proposal Algorithm}
\label{sec:wwa}

In this \add{s}ection, we introduce 
WWA which advances existing literature by (i) speeding MCMC convergence\add{,} (ii) improving the mixing of the chain \add{and} (iii) reducing comput\add{ing} time.

Algorithm~\ref{alg:dcbf} describes the double conditional Bayes factor (DCBF) sampler from \citet{Hinne2014}.
We use DCBF as a prototype for Bayesian algorithms suggested in the literature that allow for posterior inference on non-de\add{c}omposable graphs in the context of GGMs with the $G$-Wishart prior.
In the next three sections, we describe how our strategy allows us to overcome the main bottlenecks of such approaches.

\begin{algorithm}[tb]
\caption{\citep{Hinne2014} A single DCBF MCMC Step. \label{alg:dcbf}}
\textbf{Input:} Graph $G$.

\textbf{Output:} MCMC update for $G$
that preserves the posterior $p(G\mid Y)$.

For each edge $e \in {\{(i,j)\mid 1\leq i<j\leq p \}}$, do the following:
\begin{enumerate}
	\item
	Let $\widetilde{G}=(V,\widetilde{E})$ where $\widetilde{E} = E\cup\{e\}$ if $e\notin E$
	and $\widetilde{E} = E\setminus\{e\}$ otherwise.
	\item
	\label{step:dcbf_reorder}
	Reorder the nodes in $G$ and $\widetilde{G}$ so that $e$ connects node\add{s}
	$p-1$ and $p$. Rearrange $D$, $D^\star$ accordingly. \add{Denote} all resulting quantities after reordering \add{by} a superscript $e$.
	\item
	\label{step:gwish}
	Draw $K^e \mid G,Y \sim \mathcal{W}_{G^e}(\delta^\star, D^{\star,e})$ and $\widetilde{K}^{0,e} \mid \widetilde{G} \sim \mathcal{W}_{\widetilde{G}^e}(\delta, D^e)$.
	Compute their respective upper triangular Cholesky decompositions $\Phi^e$ and $\widetilde{\Phi}^{0,e}$.
	\item \label{step:accept}
	Set $G=\widetilde{G}$ w.p.~$1\wedge R_\textnormal{exchange}$ where $R_\textnormal{exchange}$ is given by Equation~\eqref{eq:r_exchange}.
	\end{enumerate}
\end{algorithm}

First, we briefly explain the derivation of the acceptance probability of the DCBF sampler.
We defer a more extensive and general explanation to Section~\ref{ap:accept_prob} of the Appendix where we derive the WWA acceptance probabilities.
Let $K^e=\left(\Phi^e\right)^\top\Phi^e$, where $\Phi^e$ is an upper triangular matrix and $K^e$ is obtained from $K$ 
by reordering the nodes such that the edge involved in the proposed graph change corresponds to nodes in the last two rows (columns) of $K^e$.
Let
$\Phi^e_{ij}$, $i,j \in \{1,\ldots,p\}$, denote the elements of the Cholesky decomposition, and define $\Phi^e_{-f} := \Phi^e\setminus\{\Phi^e_{p-1,p}, \Phi^e_{pp}\}$.
Then, consider as target the distribution $p(G,\Phi^e_{-f}\mid Y)$ as implied by the full posterior $p(G,K\mid Y) \propto p(G)\, p(K\mid G)\, p(Y\mid K)$. To compute the acceptance probability in Step~\ref{step:accept} of Algorithm~\ref{alg:dcbf}, we need
the expression
\citep[see][for a derivation]{Cheng2012}
\begin{equation}
\label{eq:exactaccratio}
    \frac{p(Y,\Phi^e_{-f}\mid \widetilde{G})}{p(Y,\Phi^e_{-f}\mid G)}
    =
	N(\Phi^e_{-f}, D^{\star,e})^{|\widetilde{E}| - |E|}\, \frac{I_G(\delta, D)}{I_{\widetilde{G}}(\delta, D)}
\end{equation}
where $N(\Phi^e_{-f}, D^{\star,e})$ is an analytically available quantity:
\begin{equation} \label{eq:N}
	N(\Phi^e_{-f}, D^{\star,e}) := \Phi^e_{p-1,p-1} \sqrt{\frac{2\pi}{D^{\star,e}_{pp}}}\,\exp\left\{ \frac{D^{\star,e}_{pp}}{2}\left(\frac{\Phi^e_{p-1,p-1} D^{\star,e}_{p-1,p}}{D_{pp}^{\star,e}} - \frac{\sum_{i=1}^{p-2} \Phi^e_{i,p-1}\Phi^e_{ip}}{\Phi^e_{p-1,p-1}} \right)^2 \right\}
\end{equation}
The ratio in \eqref{eq:exactaccratio} is not of direct use due to the intractable normalising constants $I_G(\delta, D)$, $I_{\widetilde{G}}(\delta, D)$.
The exchange algorithm \citep{Murray2006} avoids the computation of the normalising constant via the introduction of a Metropolis step defined on an augmented target distribution, which still admits as marginal the desired posterior $p(G\mid Y)$. Specifically, as shown in Step~\ref{step:gwish} \add{of} Algor\add{i}thm~\ref{alg:dcbf}, the exchange algorithm requires simulating $\widetilde{K}^{0,e} $ from the $G$-Wishart prior based on the proposed graph $\widetilde{G}$. Let 
 $\widetilde{\Phi}^{0,e}_{-f}$ be defined analogously to $\Phi^e_{-f}$
and denote its distribution
by $p(\widetilde{\Phi}^{0,e}_{-f}\mid \widetilde{G})$. Consider the distribution defined on the augmented space
\begin{align}
\label{eq:aug1}
p(G,\Phi^e_{-f},\widetilde{G},\widetilde{\Phi}^{0,e}_{-f}\mid Y)\propto 
p(G,\Phi^e_{-f}\mid Y)\, p(\widetilde{\Phi}^{0,e}_{-f}\mid\widetilde{G}).
\end{align}
DCBF proposes the deterministic exchange $G\leftrightarrow \widetilde{G}$ on the above target. Standard application of detailed balance identifies the acceptance probability as 
$1\wedge R_\textnormal{exchange}$ with
\begin{equation} \label{eq:r_exchange}
\begin{aligned}
    R_\textnormal{exchange}
    &= 
    \frac{
    p(\widetilde{G},\Phi^e_{-f},G,\widetilde{\Phi}^{0,e}_{-f}\mid Y)
    }{
    p(G,\Phi^e_{-f},\widetilde{G},\widetilde{\Phi}^{0,e}_{-f}\mid Y)
    }
    =
    \frac{p(Y,\Phi^e_{-f}\mid \widetilde{G})\, p(\widetilde{G})\, p(\widetilde{\Phi}^{0,e}_{-f}\mid G)
    }{
    p(Y,\Phi^e_{-f}\mid G)\, p(G)\, p(\widetilde{\Phi}^{0,e}_{-f}\mid\widetilde{G})}
    \\[0.4cm]
    &=  \frac{p(\widetilde{G})}{p(G)}
	\left\{ \frac{N(\Phi^e_{-f}, D^{\star,e})}{N(\widetilde{\Phi}^{0,e}_{-f}, D^e)} \right\}^{|\widetilde{E}| - |E|}
\end{aligned}
\end{equation}
where the last equality follows from \eqref{eq:exactaccratio}.

\subsection{Full Conditionals for \texorpdfstring{$K$}{K}}
\label{sec:gibbs}

In Algorithm~\ref{alg:dcbf}, sampling from the $G$-Wishart distributions in Step~\ref{step:gwish} is computationally expensive. Moreover, 
sampling from $\mathcal{W}_{G^e}(\delta^\star, D^{\star,e})$
can be considerably slower than sampling from $\mathcal{W}_{\widetilde{G}^e}(\delta, D^e)$
under the default hyperparameter choice $D = I_p$.
To avoid repeated sampling of the full matrix $K$,
WWA updates only the elements
in the Cholesky decomposition of $K$ that are affected by the change in the graph.
WWA makes use of the following conditional distributions.
First,
for $\Phi^e_{p-1,p}$,
\begin{subequations} \label{eq:phi_prop}
\begin{align}
\label{eq:a}
\Phi^e_{p-1,p} \mid
G,\,\Phi^e_{-f},\, Y &\sim \mathcal{N}\left(\frac{-\Phi^e_{p-1,p-1} D^{\star,e}_{p-1, p}}{D^{\star,e}_{p, p}}, \frac{1}{D^{\star,e}_{pp}}\right), \quad &e\in E, \\
\label{eq:b}
\Phi^e_{p-1,p} \mid
G,\,\Phi^e_{-f},\, Y &= -\frac{1}{\Phi^e_{p-1,p-1}} \sum_{l=1}^{p-2} \Phi^e_{l,p-1}\Phi^e_{lp}, \quad &e\notin E,
\end{align}
\end{subequations}
where Equation~\eqref{eq:a} follows from Equation~(8) of \citet{vandenBoom2022}
and
Equation~\eqref{eq:b} is Equation~(10) of \citet{Roverato2002}.
For $\Phi^e_{pp}$, we derive in Section~\ref{ap:deriv} of the Appendix
\begin{equation}
\label{eq:c}
D^{\star,e}_{pp}\, (\Phi^e_{pp})^2 \mid G,\,\Phi^e_{p-1,p},\,\Phi^e_{-f},
\, Y \sim \chi^2(\delta^\star).
\end{equation}
Note that the idea of updating only $\Phi^e_{p-1,p}$ and $\Phi^e_{pp}$ has already been mentioned 
in
\citet{Cheng2012} but as part of an approximate rather than \add{an} exact MCMC algorithm.

\subsection{Approximations}

We consider the approximation for the ratio of intractable normalising constants derived by \cite{Mohammadi2021}
under the default prior choice $D=I_p$. That is,
\begin{equation} \label{eq:appr}
	\frac{I_G(\delta, D)}{I_{\widetilde{G}}(\delta, D)} \,\,\approx
	\,\,\left\{\frac{\Gamma\left(\frac{\delta+d_{\widetilde{G}}}{2}
	\right)}{2\sqrt{\pi}\, \Gamma\left(\frac{\delta+d_{\widetilde{G}}+1}{2}\right)} \right\}^{|\widetilde{E}| - |E|} =: 
	\widehat{I_G/I_{\widetilde{G}}}
\end{equation}
where $d_{\widetilde{G}}$ is the number of paths of length two linking the endpoints of edge $e$.

Notice that one can avoid working with the extended target in \eqref{eq:aug1}, and apply the exchange step $G \leftrightarrow \widetilde{G}$ directly on the target $p(G,\Phi^e_{-f}\mid Y)$ with acceptance probability $1\wedge R$ where
\begin{align}
\label{eq:R}
R =  \frac{p(\widetilde{G},\Phi^e_{-f}\mid Y)
    }{
    p(G,\Phi^e_{-f}\mid Y)} \equiv \frac{p_u(\widetilde{G},\Phi^e_{-f}\mid Y)
    }{
    p_u(G,\Phi^e_{-f}\mid Y)}\times \frac{I_G(\delta, D)}{I_{\widetilde{G}}(\delta, D)}
\end{align}
for the analytically available unnormalised densities $p_u(\cdot,\cdot\mid\cdot)$ defined in the obvious way 
via~\eqref{eq:exactaccratio}. 
From \eqref{eq:appr}, \eqref{eq:R}, one can obtain the approximation
\begin{align}
\label{eq:Rhat}
\widehat{R}\equiv \widehat{R}(G,\widetilde{G},K) := \frac{p_{u}(\widetilde{G},\Phi^e_{-f}\mid Y)}{p_{u}(G,\Phi^e_{-f}\mid Y)}\times \widehat{I_G/I_{\widetilde{G}}}
\end{align}
We make use of this approximation both within the development of our informed proposal and for the introduction of a delayed acceptance step within WWA.
Combining \eqref{eq:exactaccratio}, \eqref{eq:R} leads to an explicit expression for $\widehat{R}$:
\[
	\widehat{R} =
	\frac{p(\widetilde{G})}{p(G)}
		\left\{ N(\Phi^e_{-f}, D^{\star,e})\, \frac{\Gamma\left(\frac{\delta+d_{\widetilde{G}}}{2}\right)}{2\sqrt{\pi}\, \Gamma\left(\frac{\delta+d_{\widetilde{G}}+1}{2}\right)}
	\right\}^{|\widetilde{E}| - |E|}
\]

\subsection{Informed Proposal}
\label{sec:informed}

WWA improves MCMC convergence and mixing per $G$-Wishart sample through the use of a proposal distribution that is informed by the target. We will first describe a simple modification \add{of} the proposal that is blind to the target before proceeding to the description of the informed approach.

First, notice that at every MCMC iteration, Algorithm~\ref{alg:dcbf}
scans through all the edges. At each such substep, it proposes graph $\widetilde{G}$ with the edge removed if it is present in the current graph $G$ or vice versa.
This is similar in rational\add{e} to specifying a uniform proposal distribution $q(\widetilde{G}\mid G)$ on which edge to flip.
That is,
$q(\widetilde{G}\mid G) = 1/m_{\max}$ for $\widetilde{G}\in\mathsf{nbd}(G)$
where $m_{\max} = p(p-1)/2$ denotes the maximum number of edges
and the neighbourhood $\mathsf{nbd}(G)$ is the set of $m_{\max}$ graphs that differ from $G$ by exactly one edge.
A downside of the uniform proposal is that the probability of removing an edge equals $|E|/m_{\max}$ which is usually small, especially when a shrinkage prior on graphs is used.
A possible solution is offered by \citet{Dobra2011b} who first propose to remove or add an edge with probability 0.5 and then pick an edge uniformly at random from the appropriate subset. Obviously, for
 $|E|\in\{0,\, m_{\max}\}$, we propose to add and remove an edge accordingly.
This results in the proposal
\begin{equation} \label{eq:baseQ}
q(\widetilde{G}\mid G) = \begin{cases}
	\frac{1}{m_{\max}}, &|E|=0,\, m_{\max}, \\
	\frac{1}{2|E|}, &|\widetilde{E}| < |E|\ne m_{\max}, \\
	\frac{1}{2(m_{\max} - |E|)}, &|\widetilde{E}| > |E|\ne 0, \\
\end{cases}
\end{equation}
for $\widetilde{G}\in\mathsf{nbd}(G)$.

Second, and most importantly, WWA makes use of a proposal for the graph that learns from the target.
Locally balanced proposals
\citep{Zanella2019} provide inspiration to further improve $q(\widetilde{G}\mid G)$ defined above.
Such proposals
are informed by an \textit{embarrassingly parallel} scan through the neighbourhood
of the current discrete state in a Markov chain.
Specifically,
denote the current and proposed states by $x$ and $\widetilde{x}$, respectively,
the target distribution by $\pi(x)$
and some baseline proposal by $q(\widetilde{x}\mid x)$.
Then, an informed proposal $Q(\widetilde{x}\mid x)$ is defined by
\begin{align}
\label{eq:Bal}
	Q(\widetilde{x}\mid x) \propto g\left\{\frac{\pi(\widetilde{x})}{\pi(x)}\right\}\, q(\widetilde{x}\mid x),
\end{align}
for some balancing function $g(t)$.
Here, the transition kernel $Q(\widetilde{x}\mid x)$ is locally balanced
if and only if $g(t)= t\, g(1/t)$ \citep{Zanella2019}. 

The role of the balancing function is to redirect the proposal towards candidates of high posterior probability. The aggressiveness of the redirection is determined by the shape of $g(\add{t})$. In practice, best MCMC mixing results from a balance between information from the neighbourhood scan, which concentrates the proposal, and the diffuseness of the proposal.
\add{We employ the balancing function $g(t) = t/(1+t)$
as suggested by \citet{Zanella2019}, although WWA is well defined for any $g(t)$. In our experiments,
we also consider $g(t) = \sqrt{t}$
as its unboundedness might help convergence. We find that this alternative choice results in both worse convergence and mixing (results not shown).
This reduced performance probably stems from
$g(t) = \sqrt{t}$ resulting in a too concentrated proposal.
\citet{Zanella2019}
derives
$g(t) = t/(1+t)$
as the optimal choice in an example, but also notes that
many similarly behaving balancing functions lead to virtually identical MCMC performance.
}

The use of an informed proposal in a Metropolis-Hastings acceptance probability requires the normalising constant of $Q(\widetilde{x}\mid x)$ \add{in \eqref{eq:Bal}}.
Computing the constant involves computing $\pi(\widetilde{x})/\pi(x)$
for all $\widetilde{x}$ in the support of $q(\widetilde{x}\mid x)$.
This task is \textit{embarrassingly parallel}.

WWA develops an informed proposal in the context of GGMs with $q(\widetilde{G}\mid G)$ in Equation~\eqref{eq:baseQ} as baseline proposal.
We note that the ratio $\pi(\widetilde{x})/\pi(x)$ in \eqref{eq:Bal} only serves
to improve the proposal. Instead, we use the analytically available approximation of the ratio of targets $\widehat{R}$ in \eqref{eq:Rhat}. Notice that this ratio involves the current precision matrix $K$. That is, we have 
\begin{equation} 
\label{eq:prop}
\begin{aligned}
Q(\widetilde{G}\mid G, K) &:= C(G,K)\cdot g\left\{  \frac{p_{u}(\widetilde{G},\Phi^e_{-f}\mid Y)}{p_{u}(G,\Phi^e_{-f}\mid Y)}\times \widehat{I_G/I_{\widetilde{G}}}  \right\}\,q(\widetilde{G}\mid G)\\[0.2cm]
&= C(G,K)\cdot  g\left\{\widehat{R}(G,\widetilde{G},K)\right\}q(\widetilde{G}\mid G),
\end{aligned}
\end{equation}
for a normalising constant $C(G,K)$.
Thus, the informed proposal for $\widetilde{G}$ has the form ${Q(\widetilde{G}\mid G,K)}$, differently from \cite{Zanella2019} where the informed proposal only depends on the discrete state. Moreover, in the GGM context\add{,} an update on the graph leads to an update on the precision matrix (with a distribution defined on a continuous space). Such considerations are carefully addressed in Section~\ref{ap:accept_prob} of the Appendix to ensure correctness of the deduced MCMC.

\subsection{Delayed Acceptance}

We make use of the approximation $\widehat{R}$ in \eqref{eq:Rhat} for the introduction of a delayed acceptance (DA) step \citep{Christen2005} within WWA.
We develop the DA approach by applying the idea of targeting $p(G,\Phi^e_{-f}\mid Y)$ with
a proposed exchange $G\leftrightarrow \widetilde{G}$ based on the 
kernel $Q(\widetilde{G}\mid G,K)$ in \eqref{eq:prop}, 
and a simultaneous exchange between $K\leftrightarrow \widetilde{K}$ where $\widetilde{K}$ involves the constituent elements $\widetilde{\Phi}^{e}_{-f} = \Phi^e_{-f}$, $\widetilde{\Phi}_{pp}^{e} = \Phi_{pp}^e$ (i.e., the same as the corresponding elements of $K$) and sampling $\widetilde{\Phi}^{e}_{p-1,p}$ according to \eqref{eq:phi_prop}.

Following the DA idea, 
the approximation $\widehat{R}$ in \eqref{eq:Rhat} will be used in place of the ratio 
of targets.
That is, we `promote' a proposed $\widetilde{G}$ with acceptance probability $1\wedge \widehat{R}_\textnormal{DA}$ where 
\begin{equation}
\begin{aligned}
\widehat{R}_\textnormal{DA} &:= \frac{p_{u}(\widetilde{G},\Phi^e_{-f}\mid Y)}{p_{u}(G,\Phi^e_{-f}\mid Y)}\times \widehat{I_G/I_{\widetilde{G}}}\times \frac{Q(G\mid \widetilde{G}, \widetilde{K})}{Q(\widetilde{G}\mid G, K)}  \\[0.4cm]
&= \widehat{R} \times \frac{Q(G\mid \widetilde{G}, \widetilde{K})}{Q(\widetilde{G}\mid G, K)}
\label{eq:DA}
\end{aligned}
\end{equation}
This promotion step
provides a speed-up over the exchange algorithm of acceptance probability $1\wedge R_{\textnormal{exchange}}$
by not having to sample from $\mathcal{W}_{\widetilde{G}}(\delta, D)$
at the cost
of targeting the wrong distribution. Use of the complete machinery of the
DA MCMC in the WWA algorithm corrects for this inconsistency (see Algorithm~\ref{alg:wwa}).

DA has been originally developed to employ approximate posteriors within an exact MCMC, while we directly approximate the acceptance ratio.
Specifically,
our use of DA involves first a 
Metropolis-Hastings step with approximate acceptance ratio.
Then, the outcome is treated as proposal in a second (delayed) accept-reject step that uses the exact ratio in \eqref{eq:r_exchange}.
This has the advantage that one only needs to perform the exchange algorithm when an acceptance in the approximate Metropolis-Hastings is achieved. This implies that `poor' proposed graphs get rejected quickly and more computational effort is spent on regions of the space with high posterior probability.

\begin{algorithm}
\caption{A Single WWA MCMC Step. \label{alg:wwa}}
\textbf{Input:} Graph $G$.

\textbf{Output:} MCMC update for $G$
such that the invariant distribution is the posterior $p(G\mid Y)$.
	\begin{enumerate}
		\item
		\label{step:sample_post}
		Draw ${K\mid G, Y} \sim \mathcal{W}_G(\delta^\star, D^\star)$.
		\item
		\label{step:edge_update}
		Repeat the following single-edge update a number of times $n_E$:
		\begin{enumerate}
    		\item
			\label{step:loc_bal}
			Sample $\widetilde{G}$ from the informed proposal $Q(\widetilde{G}\mid G, K)$
			given by \eqref{eq:prop}
			with
			$g(t)= t / (1+t)$.
			\item
			\label{step:reorder}
			Denote the edge in which $G$ and $\widetilde{G}$ differ by $e$.
			Reorder the nodes in $G$ and $\widetilde{G}$ so that $e$ connects node\add{s} $p-1$ and $p$. Rearrange $K$, $D$ and $D^\star$ accordingly. Denote the resulting quantities \add{by} a superscript $e$.
			\item
			\label{step:useless}
			Denote the upper triangular Cholesky decomposition of $K^e$ by $\Phi^e$.
			Update $\Phi_{pp}^e$ according to \eqref{eq:c}.
			\item
			\label{step:full_cond}
			Generate a $\widetilde{K}^{e}$ corresponding with $\widetilde{G}^e$ from $K$ by setting $\widetilde{\Phi}^{e}_{-f} = \Phi^e_{-f}$, $\widetilde{\Phi}_{pp}^{e} = \Phi_{pp}^e$, and sampling $\widetilde{\Phi}^{e}_{p-1,p}$ according to \eqref{eq:phi_prop}.
			\item \label{step:loc_bal2}
			Compute $Q(G\mid \widetilde{G}, \widetilde{K})$.
			\item
			\label{step:delayed_accept}
        	`Promote' $\widetilde{G}$ to be considered for delayed acceptance w.p.~$1\wedge \widehat{R}_\textnormal{DA}$,
			where $\widehat{R}_\textnormal{DA}$ is given by \eqref{eq:DA}.
        	If $\widetilde{G}$ is promoted:
			\begin{enumerate}
        	\item
			\label{step:sample_prior}
        	Sample $\widetilde{K}^{0,e}\mid\widetilde{G} \sim \mathcal{W}_{\widetilde{G}^e}(\delta, D^e)$.
        	\item
        	Set $G=\widetilde{G}$ and $K = \widetilde{K}$ w.p.~$1\wedge R_\textnormal{DA}$ where 
        	\[
        		R_\textnormal{DA} = R_\textnormal{exchange}\,
			\frac{(1\wedge \widehat{R}_\textnormal{DA}^{-1})\, Q(G\mid \widetilde{G}, \widetilde{K})}{(1\wedge \widehat{R}_\textnormal{DA})\, Q(\widetilde{G}\mid G,K)}
        	\]
			where $R_\textnormal{exchange}$ is given by \eqref{eq:r_exchange}.
			\end{enumerate}
		\end{enumerate}
	\end{enumerate}
\end{algorithm}

\subsection{The Complete WWA Algorithm}
\label{sec:wwa_complete}

Algorithm~\ref{alg:wwa} details the WWA algorithm.
A proposed $\widetilde{G}$ is associated with a $\widetilde{K}$
as the graph imposes a sparsity pattern on the precision matrix. To ensure detailed balance, 
 $\widetilde{K}$ appears in the computation of the reverse probability ${Q(G\mid \widetilde{G},\widetilde{K})}$.
As such,
the MCMC update needs to be joint
on $K$ and $G$ where the dimensionality $p+|E|$ of the continuous state space of $K$ varies with $G$, since every time we remove or add an edge\add{,} the number of free parameters in $K$ changes.
In Section~\ref{ap:accept_prob} of the Appendix, the resulting acceptance probability is derived via reversible jump MCMC \citep{Green1995}
to account for the transdimensionality.

In Section~\ref{ap:accept_prob} of the Appendix, we derive the acceptance probabilities involved in Step~\ref{step:delayed_accept}.
\add{The detailed balance condition on an extended space implies an acceptance probability resulting in the correct invariant distribution on the variable of interest $G$.
The construction of the extended space uses ideas from the exchange algorithm \citep{Murray2006}.
We also apply delayed acceptance 
which does not affect the invariant distribution \citep[Theorem~1]{Christen2005}
if $\widehat{R}_\textnormal{DA} > 0$,
which implies
$\widehat{I_G/I_{\widetilde{G}}} > 0$
by \eqref{eq:DA}.
Ultimately, the update is a Metropolis-Hastings step with an invariant distribution that has as marginal for $G$ the target distribution $p(G\mid Y)$.
Section~\ref{sec:iris_and_cycle} of the Appendix
confirms empirically that WWA recovers $p(G\mid Y)$.
Any approximation that satisfies $\widehat{I_G/I_{\widetilde{G}}} > 0$
leads to an MCMC that converges to $p(G\mid Y)$. For instance, 
Section~\ref{sec:convergence} of the Appendix
considers $\widehat{I_G/I_{\widetilde{G}}} = 1$
instead of Equation~\eqref{eq:appr}.
}

The relative computational cost of sampling from $\mathcal{W}_G(\delta^\star, D^\star)$ in
Step~\ref{step:sample_post}
becomes negligible
if the number of single edge updates $n_E$ is sufficiently large,
e.g.\ $n_E=p$.
Then,
the embarrassingly parallel computation of the informed proposal
in Steps~\ref{step:loc_bal} and \ref{step:loc_bal2},
and the sampling from $\mathcal{W}_{\widetilde{G}}(\delta, D)$ in Step~\ref{step:sample_prior}
carry the vast majority of computational cost (in most applications more than 90\%).

To efficiently sample from the $G$-Wishart distribution in
WWA, we combine graph decomposition with the $G$-Wishart sampler of \citet{Lenkoski2013}. 
The main idea is as follows: first, we split the graph into connected components as sampling of the rows and columns of the precision matrix can be done independently for each connected component. Note that each independent component can be sampled from a $G$-Wishart of appropriate dimension, as the entire precision matrix can be rewritten as a block matrix.
$G$-Wishart sampling for a connected component proceeds using a perfectly ordered clique minimal separator decomposition \citep[see, for example,][for an introduction to graph decomposition]{Berry2010} as detailed in \citet{Wang2010}. Note that \citet{Carvalho2007} first mention the idea of sampling from the $G$-Wishart exploiting a decomposition of the graph. We opt for the MCSM-Atom-Tree algorithm of \citet{Berry2014} to compute a perfectly ordered clique minimal separator decomposition at negligible cost.
The decomposition splits the graphs in complete (i.e.\ cliques) and incomplete prime graphs. In the first case, we can use a standard Wishart sampler, while the latter requires sampling from a $G$-Wishart.
WWA uses the $G$-Wishart sampler of \citet{Lenkoski2013} for the incomplete prime graphs.
The rejection sampler of \citet{Wang2010} is an alternative for small incomplete prime graphs where it can be faster,
but any speed-up would be negligible as the main computational cost derives from sampling large incomplete prime graphs.
\add{We empirically show in Section~\ref{sec:rgwish} of the Appendix that graph decomposition can substantially speed up sampling from the $G$-Wishart distribution
for sparse graphs,
and
that most graphs are not sparse and do not have an effective decomposition for $p\geq 20$ nodes, in which case the method from \citet{Lenkoski2013} without graph decomposition performs similarly.

The $G$-Wishart sampler of \citet{Lenkoski2013}
is an iterative algorithm initialised at a Wishart random variate. We compare this approach with initialising at a $\bar{G}$-Wishart random variate for some decomposable graph $\bar{G}$ that contains all edges of $G$, i.e.\ $E\subset\bar{E}$.
Such initialisation also results in a $G$-Wishart sampler.
However, we found that initialising with a minimal triangulation $\bar{G}$ of $G$ results in similar or increased computational cost, depending on the $G$, compared to the method from \citet{Lenkoski2013} (results not shown).
}

WWA requires a reordering of the nodes in Step~\ref{step:reorder} as explained in Section~\ref{sec:gibbs}.
In Step~\ref{step:useless}, we update $\Phi_{pp}^e$, which is not necessary for a valid MCMC as its distribution does not depend on the edge $e$ being in the graph or not.
Nonetheless, WWA includes it as its computational cost
is negligible and it improves mixing,
especially for a large number of single edge updates $n_E$.
The CL algorithm in \citet{Cheng2012} also includes the step.

\subsection{Related Work}
\label{sec:related}

In this \add{s}ection, we contextualise WWA in reference to previous work on MCMC for graphs.
\citet{Wang2012} and \citet{Cheng2012}
also consider doing a single edge update many times
for each
full update of $K$ as in Step~\ref{step:edge_update} of Algorithm~\ref{alg:wwa}.
Additionally,
they describe a two stage procedure that resembles delayed acceptance. In Section~\ref{ap:cl} of the Appendix, we describe the CL algorithm of \citet{Cheng2012}.
Its first accept-reject step (for individual edges) uses Barker's algorithm \citep{Barker1965}
with acceptance ratio $R_\textnormal{exchange}$ given in \eqref{eq:r_exchange}
where the term $N(\widetilde{\Phi}^{0,e}_{-f}, D^e)$ is set to one. To correct for this approximation, they, then, introduce a second Metropolis-Hastings accept-reject step. This combination of steps is, in fact, a delayed acceptance, but both \citet{Wang2012} and \citet{Cheng2012}
do not explicitly justify it as such.
Effectively,
they approximate the ratio of normalising constants by one
while WWA uses \eqref{eq:appr}.

WWA's computation of
the informed proposal $Q(\widetilde{G}\mid G,K)$ is embarrassingly parallel. In this respect, so is the calculation of birth and death rates in
\citet{Mohammadi2015}. Moreover,
The R package \texttt{BDgraph} \citep{Mohammadi2019}
approximates these rates using \eqref{eq:appr} by default,
which is the same approximation used for $Q(\widetilde{G}\mid G,K)$ in Step~\ref{step:loc_bal} of Algorithm~\ref{alg:wwa}.
Unlike WWA, \texttt{BDgraph} does not correct for the fact that it uses an approximation.

The embarrassingly parallel search through the neighbourhood $\mathsf{nbd}(G)$,
which constitutes $Q(\widetilde{G}\mid G,K)$,
is reminiscent of the parallel computation enabled by shotgun stochastic search \citep[SSS,][]{Hans2007}.
\citet{Jones2005} apply SSS for stochastic search on \add{the} graph space using an approximate likelihood for GGMs.
WWA similarly enables parallel computing in an MCMC framework while still using the exact likelihood.

\section{Simulation Studies}
\label{sec:empirical}

We compare WWA (Algorithm~\ref{alg:wwa}) with DCBF (Algorithm~\ref{alg:dcbf}) as DCBF can be considered  the state of the art for MCMC in GGMs with a $G$-Wishart distribution as shown by \citet{Hinne2014}.
\add{Additional comparisons in Section~\ref{sec:iris_and_cycle} of the Appendix
consider also the CL algorithm of \citet{Cheng2012}
and
\texttt{BDgraph} \citep{Mohammadi2019}
which, unlike DCBF, do not have the exact posterior $p(G\mid Y)$ as invariant distribution.
The comparisons show that the CL algorithm can provide accurate estimates of the posterior edge inclusion probabilities despite being an approximate MCMC.
The computational efficiency of WWA's exact MCMC is comparable to the approximate MCMC of the CL algorithm in a simulation with $p=100$ nodes.
\texttt{BDgraph}'s approximations have a larger effect on inclusion probability estimates.
For completeness, we mention here that
we have not included a comparison with the WL algorithm from \citet{Wang2012}
as \citet{Cheng2012}
show that their CL algorithm
substantially
outperforms the WL algorithm in terms of computational speed and accuracy of posterior approximation.
}

We choose $n_E = p$
for the number of single edge updates in Algorithm~\ref{alg:wwa}.
To bring the computational cost of Algorithm~\ref{alg:dcbf} more in line with WWA and for a fairer comparison,
we slightly modify DCBF: instead of executing the steps in Algorithm~\ref{alg:dcbf}
for all $m_{\max}$ possible edges,
we execute them for
$p$ edges drawn uniformly at random with replacement.

To measure MCMC efficiency, we use as metric
the cost of an independent sample,
\add{which} is the comput\add{ing} time required for a unit increase in the effective sample size \citep{Fang2020}:
\[
\begin{aligned}
	\text{cost of an independent sample}
	&= \frac{\text{number of MCMC steps}}{\text{effective sample size}} \times \text{cost per step} \\
	&= \text{integrated autocorrelation time} \times \text{cost per step}
\end{aligned}
\]
This captures MCMC mixing and adjusts for computational cost.
MCMC convergence can additionally be a computational bottleneck\add{,} especially if an effective initialisation is not available.
Therefore, we also discuss convergence issues in Section~\ref{sec:gene}.
The integrated autocorrelation time
is computed for the number of edges $|E|$ using the R package
\texttt{LaplacesDemon}
\citep{Statisticat2020}.
The cost of the embarrassingly parallel computation of the informed proposal
in Steps~\ref{step:loc_bal} and \ref{step:loc_bal2} of Algorithm~\ref{alg:wwa}
is assessed based on 128 CPU cores.
To make comput\add{ing} times comparable, all methods are implemented in C++ and use the same routines as much as possible,
for instance, to sample from the $G$-Wishart distribution.

\subsection{Cycle Graphs}
\label{sec:cycle}

A major improvement in computational cost \add{of}
WWA over DCBF derives from reducing the number of times we need to sample from the $G$-Wishart distribution.
The gains associated with WWA will thus be larger if $G$-Wishart sampling is slower. This situation arises, for example, when $G$ contains large incomplete prime graphs. To highlight this point, we consider 
cycle graphs, which are themselves incomplete prime graphs.
We follow Section~6.2 of \citet{Wang2012} to simulate data from cycle graphs. In the $G$-Wishart prior, we set $\delta = 3$ and $D=I_p$.
The edges are a priori independent with edge inclusion probability $\rho = 2/(p-1)$. That is,
$p(G)= \rho^{|E|}(1-\rho)^{m_{\max} - |E|}$.
We simulate $n=\tfrac{3}{2} p$ random vectors $Y_i$ from $\mathcal{N}(0_{p\times 1},\, K^{-1})$ with a precision matrix $K$ given by $K_{ii}=1$ for $i=1,\dots,p$,
$K_{ij} = 0.5$ for $|i-j|=1$, $K_{1p}=K_{p1}=0.4$ and all other elements being equal to zero.
We simulate data for $p=10,20,40$.
The performance of the two algorithms is assessed over 32 replicates of the simulations. 

We run the MCMC for 11,000 iterations, discarding the first 1,000 as burn-in. We initialise the graph at the true cycle for all algorithms. We compare the performance of WWA with (i) DCBF;
(ii) WWA without the delayed acceptance and the informed proposal;
(iii) WWA with delayed acceptance but without the informed proposal; and
(iv) WWA with the informed proposal but without delayed acceptance.
When we do not use the informed proposal, we set
$q(\widetilde{G}\mid G)$ equal to \eqref{eq:baseQ}.
When we do not perform delayed acceptance in WWA,
we use the acceptance probability in Step~\ref{step:accept} of Algorithm~\ref{alg:dcbf} directly.
These extra comparisons provide insight into the role of the different innovations of WWA.

\begin{figure}[tb]
\centering
\includegraphics[width=\textwidth]{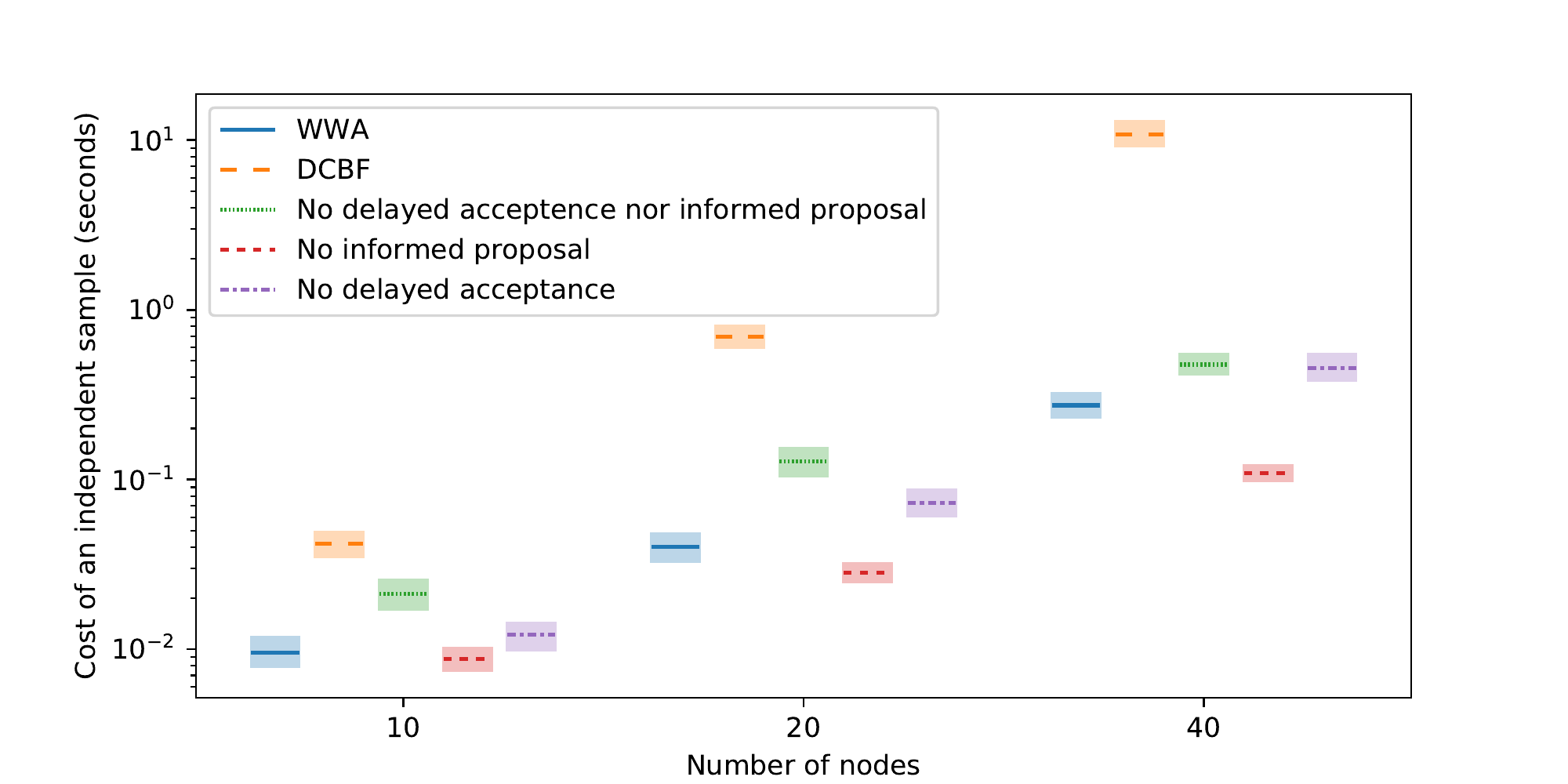}
\caption{ 
Cost of posterior computations versus the number of nodes for the data simulated from a cycle graph.
The lines represent means over the 32 repetitions for \add{DCBF} and WWA as well as different specifications of WWA.
The shaded areas are 95\% bootstrapped confidence intervals.
}
\label{fig:cycle}
\end{figure}

Figure~\ref{fig:cycle}
shows that WWA provides more efficient posterior computation than DCBF on these simulated data.
This difference increases with the number of nodes with WWA being \add{$39$} times more efficient than DCBF for $p=40$ nodes.
The MCMC without the informed proposal
outperforms WWA for $p=40$.
This is probably a result of the informed proposal and the delayed acceptance both relying on the same approximation in \eqref{eq:appr}:
the informed proposal, compared to the base proposal $q(\widetilde{G}\mid G)$, increases the acceptance ratio of the first approximate accept-reject step in DA MCMC,
but this increased acceptance does not translate to a proportional increase in the overall acceptance ratio.
These effects are compounded by
the approximation in \eqref{eq:appr} becoming less accurate for larger graphs \citep{Letac2007}.
The result is that
Step~\ref{step:sample_prior} of Algorithm~\ref{alg:wwa},
which
involves the relatively expensive sampling from $\mathcal{W}_{\widetilde{G}}(\delta, D)$,
is evaluated more often without a corresponding improvement in MCMC mixing, i.e.\ the gain in mixing from the informed proposal does not compensate for this extra sampling.

\subsection{Uniformly Sampled Graphs}
\label{sec:uniform}

In this \add{s}ection, we compare the performance of the different algorithms on data simulated from the Bayesian model
described in Section~\ref{sec:model}
with $n = 2p$ and a uniform prior on graphs: $p(G) = 2^{-m_{\max}}$. In particular, \add{for each replicate,} we generate \add{a graph $G$ from} this uniform distribution \add{$p(G)$, sample a precision matrix $K$ from $\mathcal{W}_G(\delta,D) = \mathcal{W}_G(3,I_p)$ and data from $\mathcal{N}(0_{p\times 1},\, K^{-1})$}.
We show results for 32 replicates and
for $p=10,20,40$.
MCMC is initialised at the true graph $G$.
The remaining set-up of this simulation study follows Section~\ref{sec:cycle}.

\begin{figure}[tb]
\centering
\includegraphics[width=\textwidth]{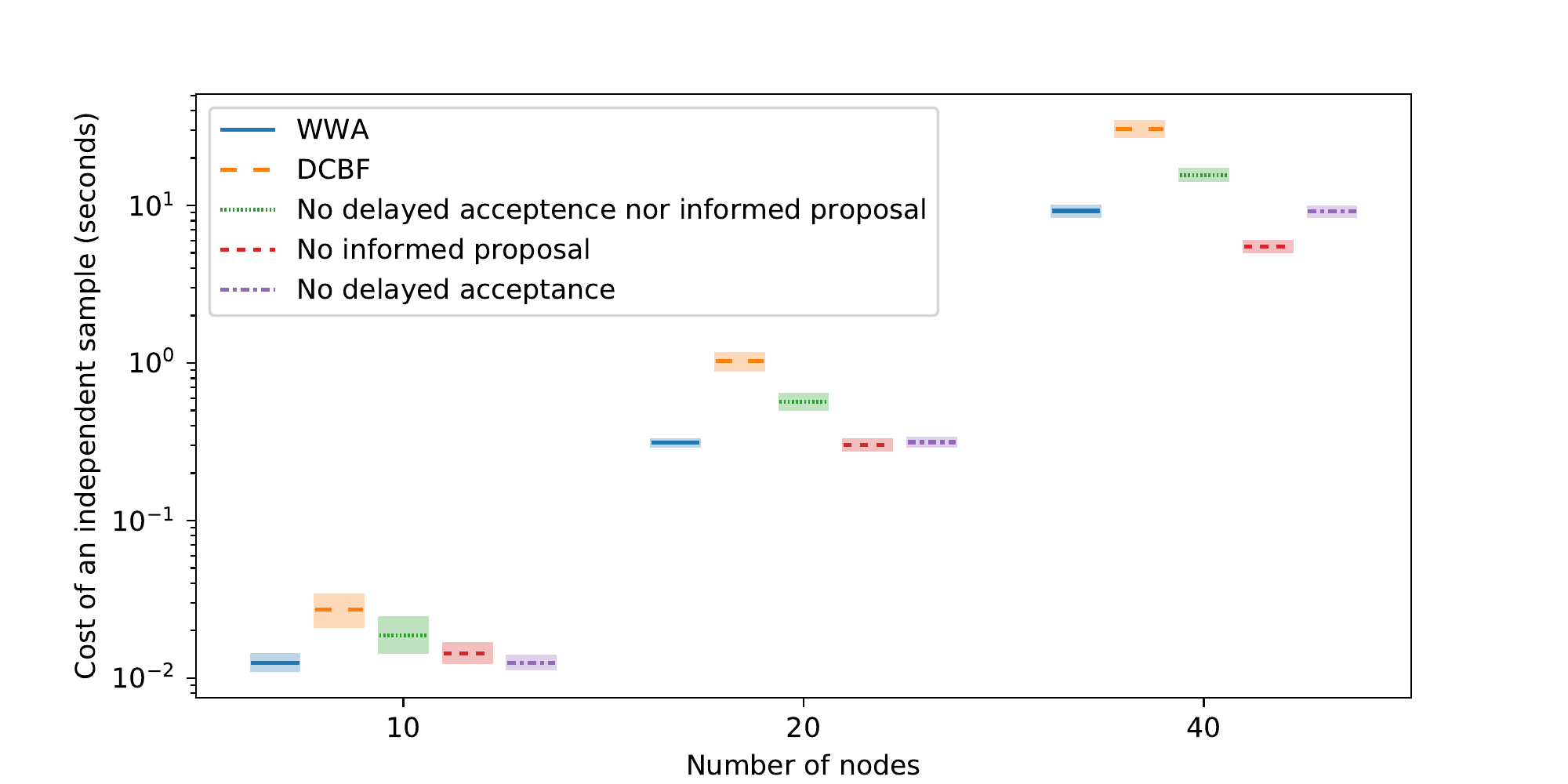}
\caption{
Cost of posterior computations versus the number of nodes for the data simulated from a uniformly sampled graph.
The lines represent means over the 32 repetitions for \add{DCBF} and WWA as well as different specifications of WWA.
The shaded areas are 95\% bootstrapped confidence intervals.
}
\label{fig:simul}
\end{figure}

Also in this simulation scenario, WWA provides more efficient posterior computation than DCBF as shown in Figure~\ref{fig:simul}.
Again, this difference increases with the number of nodes with WWA being $3.3$ times more efficient than DCBF for $p=40$ nodes.
As in Section~\ref{sec:cycle}, the MCMC without the informed proposal
outperforms WWA for $p=40$.

\section{Application to Gene Expression Data}
\label{sec:gene}

We consider the real data application from Section~4.2 of \citet{Mohammadi2015} where a more extensive data description is available.
The data consist of gene expressions in B-lymphocyte cells \citep{Stranger2007} from $n=60$ individuals.
They
are quantile-normalised to marginally follow a standard Gaussian distribution, a process also known as rank normalisation.
We consider two data sets $Y$, namely those consisting of the $p=50$ and $p=100$ most variable gene expressions.
The prior on $(G,K)$ is the same as in Section~\ref{sec:uniform}, which coincides with an uni\add{n}formative prior.

For $p=50$,
WWA and DCBF are initialised at a graphical lasso estimate of the graph~$G$ \citep{Friedman2007} and
are run for 16,000 iterations of which the last 10,000 are used to estimate the cost of an independent sample.
For the data set with $p=100$,
the number of possible graphs
is $2^{m_{\max}} = 1.3\cdot 10^{1,490}$, and the precision matrix is not identifiable in the likelihood since $n<p$. Although this is \add{theoretically} not a problem in the Bayesian framework because of prior regularisation, likelihood unidentifiability is known to cause problems for MCMC convergence.
Also, the approximation in \eqref{eq:appr} favours sparse graphs
as it is consistently biased in this direction \citep[page~13]{Mohammadi2021},
a tendency which is rather strong when the posterior is not concentrated as in the case of $p=100$.
As a result, using the informed proposal or delayed acceptance based on \eqref{eq:appr} results in bad MCMC convergence and mixing.
Therefore, we use 
Algorithm~\ref{alg:wwa}
without the informed proposal nor delayed acceptance because the bias in the approximation would dominate the information deriving from the posterior which is flat in this example.

\begin{figure}
\centering
\includegraphics[width=\textwidth]{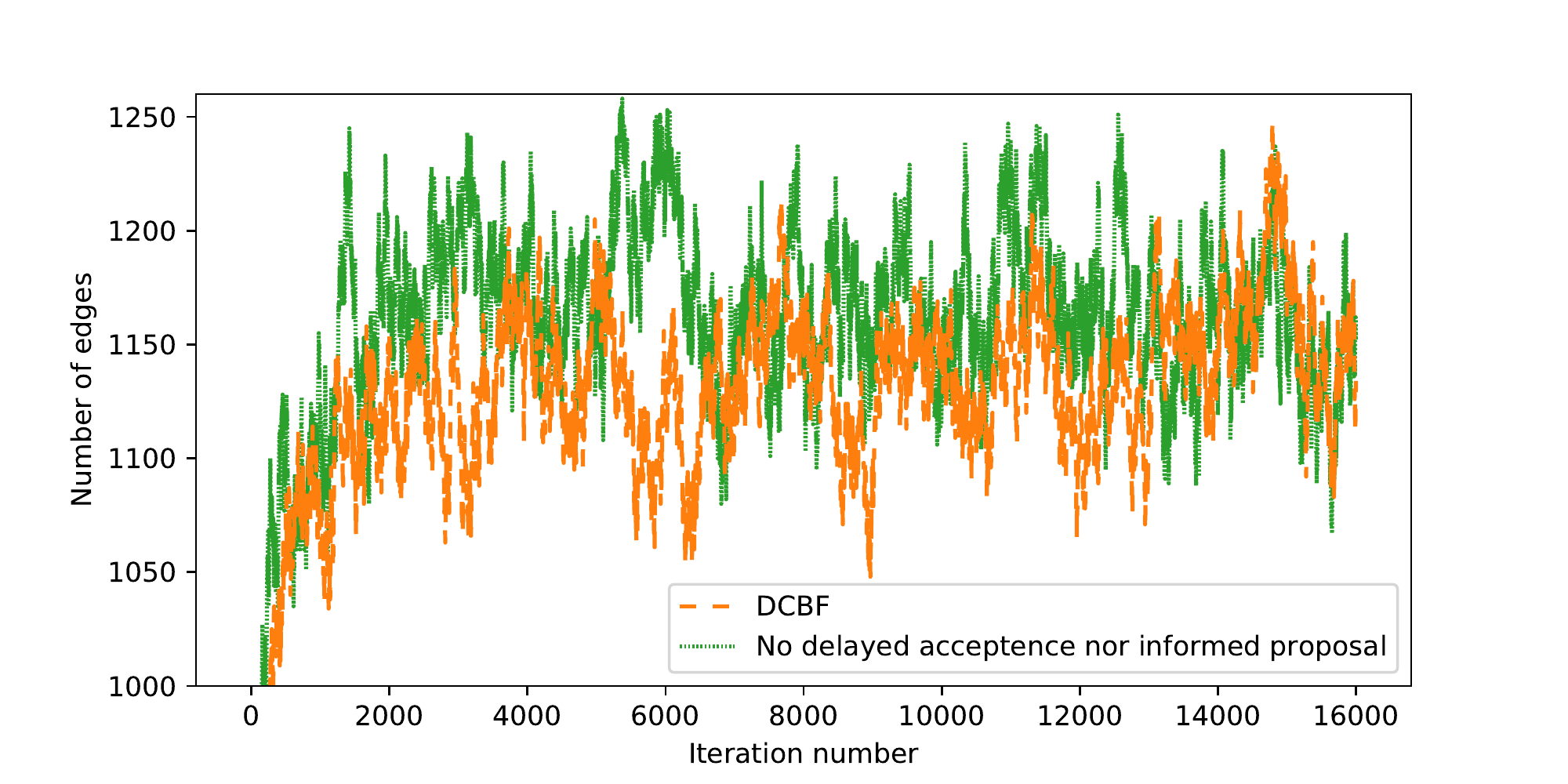}
\caption{
Trace plots for the number of edges in the gene expression application with $p=100$ nodes.
}
\label{fig:gene}
\end{figure}

In terms of speed, the first 6,000 burn-in iterations take 11 minutes for WWA versus 36 minutes for DCBF with $p=50$, and 4.1 hours for the proposed algorithm versus 9.0 hours for DCBF with $p=100$.
We compute the improved $\widehat{R}$ of \citet{Vehtari2021} as a diagnostic of convergence on the last 10,000 iterations. $\widehat{R}$ converges to one as the number of iterations tends to infinity and $\widehat{R} > 1.01$ indicates lack of convergence.
For $p=50$, when the precision matrix is likelihood identifiable as well, convergence is reached quickly by both algorithms with $\widehat{R}=1.003$ for WWA and $\widehat{R}=1.006$ for DCBF.
For $p=100$,
the trace plots in
Figure~\ref{fig:gene}
show an increasing trend for the number of edges for DCBF,
indicating slower convergence confirmed by $\widehat{R}=1.068$.
This contrasts with the proposed algorithm where $\widehat{R}=1.001$ and which seems to converge within 2,000 iterations.
The introduced methodology
yields faster MCMC convergence,
both in terms of number of iterations and especially in terms of comput\add{ing} time.

WWA is also superior to DCBF in terms of MCMC mixing.
For $p=100$, we run DCBF for 10,000 iterations initialised at the last iteration of the proposed algorithm
to avoid DCBF's convergence issues while assessing MCMC mixing. The cost of an independent sample is
6.2 seconds for WWA
versus
24 seconds for DCBF
with $p=50$,
and
7.4 minutes for the proposed algorithm
versus
14
minutes for DCBF
with $p = 100$.

\section{Discussion}
\label{sec:discussion}

In this work, we propose WWA, a novel algorithm for GGMs which significantly improves \add{on} existing MCMC based on the $G$-Wishart prior.
The main contributions involve delayed acceptance, Gibbs updates for the precision matrix, and the use of parallel computing to inform the proposal.
As a result, WWA outperforms the state-of-the-art alternative in terms of MCMC mixing, convergence and comput\add{ing} time.

Here, we discuss possible improvements and extensions to WWA.
As the number of nodes increases,
normalising the informed proposal takes longer,
with the added computational cost potentially outweighing the improvement in MCMC mixing.
A potential extension to tune this computation versus mixing
trade-off is blocking \citep{Zanella2019}.
It constrains the support of the informed proposal to a subset or `block' of the neighbourhood of a graph, reducing \add{the} computational cost of normalising the informed proposal.

Another improvement to the informed proposal would be a faster or more accurate approximation than \eqref{eq:Rhat}.
The computational bottleneck of the current approximation is calculating the Cholesky decomposition $\Phi^e$.
For sparse graphs,
the
Cholesky decomposition
can be sped up via fill-in reducing node reorderings, which increase sparsity in $\Phi^e$, 
and Cholesky routines optimised for sparse matrices \citep[Section~2.4.3]{Rue2001,Rue2005}
as shown by
\citet{Cheng2012}.

The MCMC performance is limited by the fact that at most one edge is changed for each accept-reject step.
A truly scalable algorithm requires larger moves in the graph space as the number of possible edges in a graph is quadratic in the number of nodes.
Such larger moves require sufficiently good proposals for both the graph and the precision matrix $K$.
\citet{Tan2017} take a first step in this direction by changing multiple edges at the same time with an approximate likelihood on the graph resulting from approximating $I_G(\delta, D)$.

In this work, we focus on GGMs because of their popularity in applications and the computational challenges associated with their estimation. Due to the modular nature of MCMC, WWA can also provide a feasible strategy in extended models such as multiple graphs \citep[e.g.,][]{Peterson2015,Tan2017}, Gaussian copulas to accommodate non-Gaussian data \citep[e.g.,][]{Dobra2011a,Mohammadi2019} or sparse seemingly unrelated regressions \citep[e.g.,][]{Wang2010b,Bhadra2013}.

\bigskip
\begin{center}
{\large\bf SUPPLEMENTARY MATERIAL}
\end{center}

\begin{description}

\add{\item[Appendix:] Derivations of Equation~\eqref{eq:c} and WWA's acceptance probabilities, description of the algorithm from \citet{Cheng2012},
and additional empirical results. (.pdf file)}

\item[Code:] The scripts that produced the empirical results are available at \url{https://github.com/willemvandenboom/wwa}. (GitHub repository)

\end{description}

\bibliographystyle{chicago}
\bibliography{graph}

\end{document}


\def\spacingset#1{\renewcommand{\baselinestretch}%
{#1}\small\normalsize} \spacingset{1}


\if0\blind
{
  \title{\bf Appendix to ``The $G$-Wishart Weighted Proposal Algorithm:
Efficient Posterior Computation for Gaussian Graphical Models'' published in the Journal of Computational and Graphical Statistics}
  \author{Willem van den Boom, Alexandros Beskos and Maria De Iorio}
  \maketitle
} \fi

\spacingset{1.5} 

\spacingset{1.5} 

\section{Derivation of Equation~\texorpdfstring{\eqref{eq:c}}{(6)}}  
\label{ap:deriv}

By Equation~(20) of \citet{Roverato2002}, \\  
$p(\Phi^e\mid G,Y)\propto
    \exp\left[-\tfrac{1}{2} \mathrm{tr}\{(\Phi^e)^\top \Phi^e D^{\star,e}\}\right]
    \prod_{i=1}^p (\Phi^e_{ii})^{\delta^\star + \nu_i^{G^e} - 1}$
where \\  
$\nu_i^{G^e} = {|\{j\mid (i,j)\in E^e\ \text{and}\ i < j\}|}$.
This density factorises as  \\  
$
p(\Phi^e\mid G,Y)\propto f(\Phi^e) \exp\left\{
    -\tfrac{1}{2} (\Phi^e_{pp})^2 D^{\star,e}_{pp}
    \right\}
    (\Phi^e_{pp})^{\delta^\star - 1}
$
where $f(\Phi^e)$ is constant with respect to $\Phi^e_{pp}$.
Thus,
$p(\Phi^e_{pp}\mid G,\Phi^e_{p-1,p},\Phi^e_{-f},Y) \propto
    \exp\left\{
    -\tfrac{1}{2} (\Phi^e_{pp})^2 D^{\star,e}_{pp}
    \right\}
    (\Phi^e_{pp})^{\delta^\star - 1}$.
The transformation of variables $\Phi^e_{pp}\to (\Phi^e_{pp})^2$
yields  \\  
$p\{(\Phi^e_{pp})^2\mid G,\Phi^e_{p-1,p},\Phi^e_{-f},Y\}
\propto \exp\left\{
    -\tfrac{1}{2} (\Phi^e_{pp})^2 D^{\star,e}_{pp}
    \right\}
    \{(\Phi^e_{pp})^2\}^{\delta^\star/2 - 1}$
from which Equation~\eqref{eq:c}
follows.

\section{Derivation of WWA's Acceptance Probabilities}
\label{ap:accept_prob}

This \add{section} derives in detail the acceptance ratios $\widehat{R}_\textnormal{DA}$ and $R_\textnormal{DA}$ used in Algorithm~\ref{alg:wwa}.
The informed proposal $Q(\widetilde{G}\mid G,K)$
depends on the value of $\Phi^e_{p-1,p}$.
Thus, we need to treat the single edge update as a joint update of $\Phi^e_{p-1,p}$ and $G$
to derive its acceptance probability.
The update is transdimensional since $\Phi^e_{p-1,p}$ is fixed or free depending on $G$ per \eqref{eq:phi_prop}.
We therefore consider reversible jump MCMC \citep{Green1995} for this derivation.
Ultimately, the exchange algorithm \citep{Murray2006}, used to avoid intractable normalis\add{ing} constants,
circumvents the need for a reversible jump
as exchanging variables does not affect their joint dimensionality,
but the intermediate reversible jump acceptance probability is used to derive the approximate acceptance ratio $\widehat{R}_\textnormal{DA}$ for use with delayed acceptance MCMC.

\subsection{Reversible Jump}

Recall $\Phi^e_{-f} = \Phi^e\setminus\{\Phi^e_{p-1,p}, \Phi^e_{pp}\}$.
By construction, $\widetilde{\Phi}^{e}_{-f}=\Phi^e_{-f}$.
WWA uses the joint proposal
\[
	q(\widetilde{G},\widetilde{\Phi}^e_{p-1,p}\mid G, K)
	= Q(\widetilde{G}\mid G,K)\,q(\widetilde{\Phi}^e_{p-1,p}\mid \widetilde{G},\Phi^e_{-f})
\]
Algorithm~\ref{alg:wwa} chooses
${q(\widetilde{\Phi}^e_{p-1,p}\mid \widetilde{G},\Phi^e_{-f})} = {p(\widetilde{\Phi}^e_{p-1,p}\mid \widetilde{G}, \widetilde{\Phi}^e_{-f}, Y)}$ given by \eqref{eq:phi_prop}
which is thus a Dirac delta function if $e\notin\widetilde{G}$.
Therefore, the previous display reduces to
\begin{equation} \label{eq:prop_rj}
	q(\widetilde{G},\widetilde{\Phi}^e_{p-1,p}\mid G, K)
	=
	\begin{cases}
	Q(\widetilde{G}\mid G,K),\quad\hfill \widetilde{\Phi}^e_{p-1,p} = \phi^e\ \text{and}\ e\notin\widetilde{E}, \\
	p(\widetilde{\Phi}^e_{p-1,p}\mid \widetilde{G}, \widetilde{\Phi}^e_{-f}, Y)\, Q(\widetilde{G}\mid G,K),\quad\hfill e\in\widetilde{E},
	\end{cases}
\end{equation}
where $\phi^e$ equals the right-hand side of \eqref{eq:b},
$Q(\widetilde{G}\mid G,K)$ is a probability mass function
and $p(\widetilde{\Phi}^e_{p-1,p}\mid \widetilde{G}, \widetilde{\Phi}^e_{-f}, Y)$ is a density with respect to Lebesgue measure.

The target distribution
is
\begin{equation} \label{eq:target_rj}
	p(G, \Phi^e_{p-1,p}\mid \Phi^e_{-f},Y)
	= \begin{cases}
		p(G\mid \Phi^e_{-f},Y),\quad\hfill \Phi^e_{p-1,p} = \phi^e\ \text{and}\ e\notin E, \\
	p(\Phi^e_{p-1,p}\mid G, \Phi^e_{-f}, Y)\, p(G\mid \Phi^e_{-f},Y),\quad\hfill e\in E.
	\end{cases}
\end{equation}
This target, including
$\Phi^e_{p-1,p}$ and $\Phi^e_{-f}$, depends on $e$ and thus on the proposed $\widetilde{G}$.
Such dependence on the proposed graph does not affect
the MCMC's validity or the acceptance probability
as the resulting Markov kernel can be interpreted as a mixture of kernels which each have $p(G,K\mid Y)$ as invariant distribution \citep[Section~2.4]{Tierney1994}.

Reversible jump involves a dimension-matching map.
In our setting, we consider the mapping $(\Phi^e_{p-1,p},\, \widetilde{\Phi}^e_{p-1,p}) \to (\widetilde{\Phi}^e_{p-1,p},\, \Phi^e_{p-1,p})$
which has matched dimensions since both sides contain exactly one free element.
Moreover, the absolute value of the determinant of the Jacobian of this map equals one.
Therefore, the acceptance probability follows as
$1\wedge R_\textnormal{RJ}$ where \citep[Equation~(7)]{Green1995}
\begin{equation} \label{eq:r_rj}
	R_\textnormal{RJ} = \frac{p(\widetilde{G}, \widetilde{\Phi}^e_{p-1,p}\mid \widetilde{\Phi}^e_{-f},Y)\, q(G,\Phi^e_{p-1,p}\mid \widetilde{G}, \widetilde{K})}{p(G, \Phi^e_{p-1,p}\mid \Phi^e_{-f},Y)\, q(\widetilde{G},\widetilde{\Phi}^e_{p-1,p}\mid G, K)}.
\end{equation}
Bayes' rule and $\widetilde{\Phi}^e_{-f}=\Phi^e_{-f}$ yields
\[
	\frac{p(\widetilde{G}\mid \widetilde{\Phi}^e_{-f},Y)}{p(G\mid \Phi^e_{-f},Y)}
	= \frac{p(Y, \widetilde{G}, \widetilde{\Phi}^e_{-f})}{p(Y, G, \Phi^e_{-f})}
	= \frac{p(Y, \widetilde{\Phi}^e_{-f}\mid \widetilde{G})\, p(\widetilde{G})}{p(Y, \Phi^e_{-f}\mid G)\, p(G)}.
\]
Combining the last two displays, and inserting
Equations~\eqref{eq:prop_rj} and \eqref{eq:target_rj}
provide
\begin{equation} \label{eq:target_rj2}
\begin{aligned}
	R_\textnormal{RJ}
	&= \frac{p(\widetilde{G},\Phi^e_{-f}\mid Y)\, Q(G\mid \widetilde{G}, \widetilde{K})}{p(G, \Phi^e_{-f}\mid Y)\, Q(\widetilde{G}\mid G,K)}
	= \frac{p(Y,\Phi^e_{-f}\mid \widetilde{G})\, p(\widetilde{G})\, Q(G\mid \widetilde{G}, \widetilde{K})}{p(Y, \Phi^e_{-f}\mid G)\, p(G)\, Q(\widetilde{G}\mid G,K)} \\
	&= N(\Phi^e_{-f}, D^{\star,e})^{|\widetilde{E}| - |E|}\, \frac{p(\widetilde{G})\, I_G(\delta, D)\, Q(G\mid \widetilde{G}, \widetilde{K})}{p(G)\, I_{\widetilde{G}}(\delta, D)\, Q(\widetilde{G}\mid G,K)},
\end{aligned}
\end{equation}
where the last equality follows from Equation~\eqref{eq:exactaccratio}.

\subsection{Exchange Algorithm}

Equation~\eqref{eq:target_rj2} cannot be used directly as it contains
the intractable $I_{G}(\delta, D)$.
Using the exchange algorithm avoids evaluating this normalis\add{ing} constant
as in \citet[Section~5.2]{Wang2012} and \citet[Section~2.3]{Cheng2012}
by considering
an augmented target distribution.

\begin{figure}
\vspace{-0.1cm}
\centering
\begin{tikzpicture}[
            > = stealth, 
            shorten > = 1pt, 
            auto,
            node distance = 3cm, 
            semithick 
        ]

        \tikzstyle{every state}=[
            draw = black,
            thick,
            fill = white,
            minimum size = 8mm
        ]

        \node[state] (G) {$G$};
        \node[state] (K) [above left of=G] {$K$};
        \node[state] (G_tilde) [above right of=K] {$\widetilde{G}$};
        \node[state] (Phi_0) [above right of=G] {$\widetilde{\Phi}^{0,e}_{-f}$};
        \node[state] (Y) [left of=K] {$Y$};

        \path[->] (G) edge node {} (G_tilde);
        \path[->] (G) edge node {} (K);
        \path[->] (G_tilde) edge node {} (Phi_0);
        \path[->] (K) edge node {} (G_tilde);
        \path[->] (K) edge node {} (Y);
        
        \path[->] (G) edge node {} (Phi_0);
    \end{tikzpicture}
\caption{
A directed acyclic graph representing
the conditional dependence structure of
the augmented target distribution \eqref{eq:augmented_target} used with the exchange algorithm.
}
\label{fig:dag}
\end{figure}
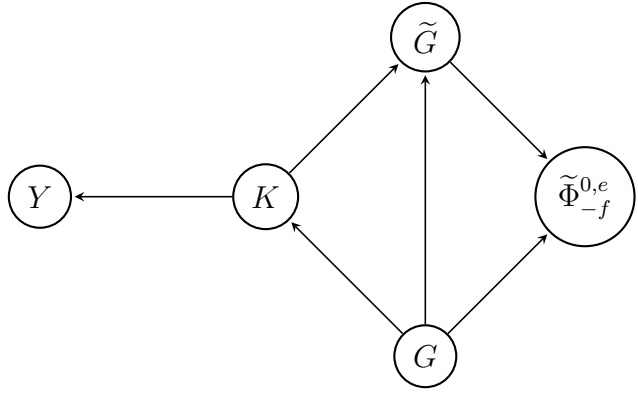

Our augmented target is a joint distribution on
$(G,K,\widetilde{G}, \widetilde{\Phi}^{0,e}_{-f})$
that mimics Algorithm~\ref{alg:wwa}.
Specifically,
$(G,K)$ follows the unaugmented target distribution $p(G,K\mid Y)$,
the distribution of $\widetilde{G}\mid G,K$ is
the informed proposal $Q(\widetilde{G}\mid G,K)$,
and the distribution of
$\widetilde{\Phi}^{0,e}_{-f}\mid G,K, \widetilde{G}$
follows from the node reordering resulting from the pair $(G,\widetilde{G})$,
$\widetilde{K}^{0,e}\mid \widetilde{G} \sim \mathcal{W}_{\widetilde{G}^e}(\delta, D^e)$
and the definition $\widetilde{\Phi}^{0,e}_{-f} = \widetilde{\Phi}^{0,e}\setminus\{\widetilde{\Phi}^{0,e}_{p-1,p}, \widetilde{\Phi}^{0,e}_{pp}\}$.
The augmented target distribution can thus be written as
\begin{equation} \label{eq:augmented_target}
    \pi(G,K,\widetilde{G}, \widetilde{\Phi}^{0,e}_{-f}) =
    p(G,K\mid Y)\, Q(\widetilde{G}\mid G,K)\,
    p(\widetilde{\Phi}^{0,e}_{-f}\mid \widetilde{G}).
\end{equation}
Figure~\ref{fig:dag} clarifies the structure of the augmented target.

The proposal is to exchange
$G$
with
$\widetilde{G}$
and
update $K$ by a value sampled from
its proposal
$q(\widetilde{K}\mid G,K,\widetilde{G})$,
which is defined by sampling $\widetilde{\Phi}^e_{p-1,p}$ according to \eqref{eq:phi_prop} while leaving the other elements of $\widetilde{\Phi}^e$
equal to the corresponding elements in $\Phi^e$.
The resulting
Metropolis-Hastings step
has the posterior of interest $p(G,K\mid Y)$ as target marginally on $(G,K)$.
The remainder of this section evaluates the corresponding acceptance probability $1\wedge R_\textnormal{exchange}^\textnormal{informed}$
with
\begin{equation} \label{eq:lb_accept_prob}
\begin{aligned}
    R_\textnormal{exchange}^\textnormal{informed}
    &= \frac{
    \pi(\widetilde{G},\widetilde{K},G, \widetilde{\Phi}^{0,e}_{-f})\,
    q(K\mid \widetilde{G},\widetilde{K}, G)
    }{
    \pi(G,K,\widetilde{G}, \widetilde{\Phi}^{0,e}_{-f})\,
    q(\widetilde{K}\mid G,K,\widetilde{G})} \\[0.2cm]
    &= \frac{
    p(\widetilde{G},\widetilde{K}\mid Y)\,
    Q(G\mid \widetilde{G},\widetilde{K})\,
    p(\widetilde{\Phi}^{0,e}_{-f}\mid G)\,
    q(K\mid \widetilde{G},\widetilde{K}, G)
    }{
    p(G,K\mid Y)\,
    Q(\widetilde{G}\mid G,K)\,
    p(\widetilde{\Phi}^{0,e}_{-f}\mid \widetilde{G})\,
    q(\widetilde{K}\mid G,K,\widetilde{G})}
\end{aligned}
\end{equation}
where the last equality follows from \eqref{eq:augmented_target}.

The distribution ${q(\widetilde{K}\mid G,K,\widetilde{G})}$
is defined through
$\widetilde{\Phi}^e_{p-1,p}$ being distributed according to \eqref{eq:phi_prop}.
The implied distribution on $\widetilde{K}$ follows from the transformation
$\widetilde{\Phi}^e \to \widetilde{K}$
such that
\[
    {q(\widetilde{K}\mid G,K,\widetilde{G})} = J(\widetilde{\Phi}^e \to \widetilde{K})\, {q(\widetilde{\Phi}^e_{p-1,p}\mid \widetilde{G},\Phi^e_{-f})}
\]
where
$J(\widetilde{\Phi}^e \to \widetilde{K})$
is the Jacobian term resulting from the transformation of variables.
Similarly,
\[
    p(G,K\mid Y) = J(\Phi^e \to K)\, p(G, \Phi^e\mid Y).
\]
These Jacobian terms cancel in
$R_\textnormal{exchange}^\textnormal{informed}$.
Specifically,
inserting
the previous two displays
into \eqref{eq:lb_accept_prob} yields
\[
    R_\textnormal{exchange}^\textnormal{informed}
    = \frac{
    p(\widetilde{G}, \widetilde{\Phi}^e\mid Y)\,
    Q(G\mid \widetilde{G},\widetilde{K})\,
    p(\widetilde{\Phi}^{0,e}_{-f}\mid G)\,
    q(\Phi^e_{p-1,p}\mid G, \widetilde{\Phi}^e_{-f})
    }{
    p(G, \Phi^e\mid Y)\,
    Q(\widetilde{G}\mid G,K)\,
    p(\widetilde{\Phi}^{0,e}_{-f}\mid \widetilde{G})\,
    q(\widetilde{\Phi}^e_{p-1,p}\mid \widetilde{G},\Phi^e_{-f})}.\vspace{0.2cm}
\]
Consider the factorisation
$p(G, \Phi^e\mid Y) =
    p(G, \Phi^e_{p-1,p}\mid \Phi^e_{pp},\Phi^e_{-f},Y)\,
    p(\Phi^e_{pp},\Phi^e_{-f}\mid Y)$.
The proposal is such that
$\widetilde{\Phi}^e_{pp} = \Phi^e_{pp}$ and $\widetilde{\Phi}^e_{-f} = \Phi^e_{-f}$.
We also make use of the equality $p(G, \Phi^e_{p-1,p}\mid \Phi^e_{pp},\Phi^e_{-f},Y) = {p(G, \Phi^e_{p-1,p}\mid \Phi^e_{-f},Y)}$
since $\Phi^e_{pp}$ is independent of $(G,\Phi^e_{-f},\Phi^e_{p-1,p})$ by \eqref{eq:c}.
Thus,
\[
    \frac{
    p(\widetilde{G}, \widetilde{\Phi}^e\mid Y)
    }{
    p(G, \Phi^e\mid Y)}
    =
    \frac{
    p(\widetilde{G}, \widetilde{\Phi}^e_{p-1,p}\mid \widetilde{\Phi}^e_{pp},\widetilde{\Phi}^e_{-f},Y)\,
    p(\widetilde{\Phi}^e_{pp},\widetilde{\Phi}^e_{-f}\mid Y)
    }{
    p(G, \Phi^e_{p-1,p}\mid \Phi^e_{pp},\Phi^e_{-f},Y)\,
    p(\Phi^e_{pp},\Phi^e_{-f}\mid Y)
    }
    = \frac{
        p(\widetilde{G}, \widetilde{\Phi}^e_{p-1,p}\mid \widetilde{\Phi}^e_{-f},Y)
    }{
        p(G, \Phi^e_{p-1,p}\mid \Phi^e_{-f},Y)
    }.
\]
Furthermore,
$Q(\widetilde{G}\mid G,K)\, q(\widetilde{\Phi}^e_{p-1,p}\mid \widetilde{G},\Phi^e_{-f})
= q(\widetilde{G},\widetilde{\Phi}^e_{p-1,p}\mid G, K)$
where ${q(\widetilde{G},\widetilde{\Phi}^e_{p-1,p}\mid G, K)}$
is given by \eqref{eq:prop_rj}.
Combined with the previous two displays,
we obtain
\[
\begin{aligned}
    R_\textnormal{exchange}^\textnormal{informed}
    &=
    \frac{
        p(\widetilde{G}, \widetilde{\Phi}^e_{p-1,p}\mid \widetilde{\Phi}^e_{-f},Y)\,
        q(G,\Phi^e_{p-1,p}\mid \widetilde{G}, \widetilde{K})\,
    p(\widetilde{\Phi}^{0,e}_{-f}\mid G)
    }{
        p(G, \Phi^e_{p-1,p}\mid \Phi^e_{-f},Y)\,
        q(\widetilde{G},\widetilde{\Phi}^e_{p-1,p}\mid G, K)\,
    p(\widetilde{\Phi}^{0,e}_{-f}\mid \widetilde{G})
    } \\
    &=
    R_\textnormal{RJ}\,
    \frac{
    p(\widetilde{\Phi}^{0,e}_{-f}\mid G)
    }{
    p(\widetilde{\Phi}^{0,e}_{-f}\mid \widetilde{G})
    }
\end{aligned}
\]
where the last equality follows from \eqref{eq:r_rj}.
Inserting \eqref{eq:target_rj2} and recalling \eqref{eq:r_exchange}
yields
\[
    R_\textnormal{exchange}^\textnormal{informed} = R_\textnormal{exchange}\, \frac{Q(G\mid \widetilde{G}, \widetilde{K})}{Q(\widetilde{G}\mid G,K)}.
\]
The acceptance probabilities in Algorithm~\ref{alg:wwa} follow now from the delayed acceptance method in Algorithm~1 of \citet{Christen2005}.
Specifically, $\widehat{R}_\textnormal{DA}$ equals $R_\textnormal{RJ}$ with the ratio of normalising constants in \eqref{eq:target_rj2} approximated by \eqref{eq:appr}.
Then, the second accept-reject step uses $R_\textnormal{exchange}$ from Equation~\eqref{eq:r_exchange} in place of the ratio of target distributions
like $R_\textnormal{exchange}^\textnormal{informed}$ does.

\add{
\section{Effect of \texorpdfstring{$\widehat{I_G/I_{\widetilde{G}}}$}{} Approximation on Convergence}
\label{sec:convergence}

WWA converges to $p(G\mid Y)$
for any approximation $\widehat{I_G/I_{\widetilde{G}}} > 0$
as remarked in Section~\ref{sec:wwa_complete}.
Here,
we investigate in a simulation study the effect
of the choice of $\widehat{I_G/I_{\widetilde{G}}}$
on the speed of convergence for WWA.
The data are simulated as in Section~\ref{sec:cycle}
for $p = 40$.
Then,
we run WWA for
1,000
iterations
both for $\widehat{I_G/I_{\widetilde{G}}}$
as given by \eqref{eq:appr}
following the suggestion of \citet{Mohammadi2021}
and for $\widehat{I_G/I_{\widetilde{G}}} = 1$.
The MCMC chains are initialised at the empty graph.

\begin{figure}[tbp]
\centering
\includegraphics[width=\textwidth]{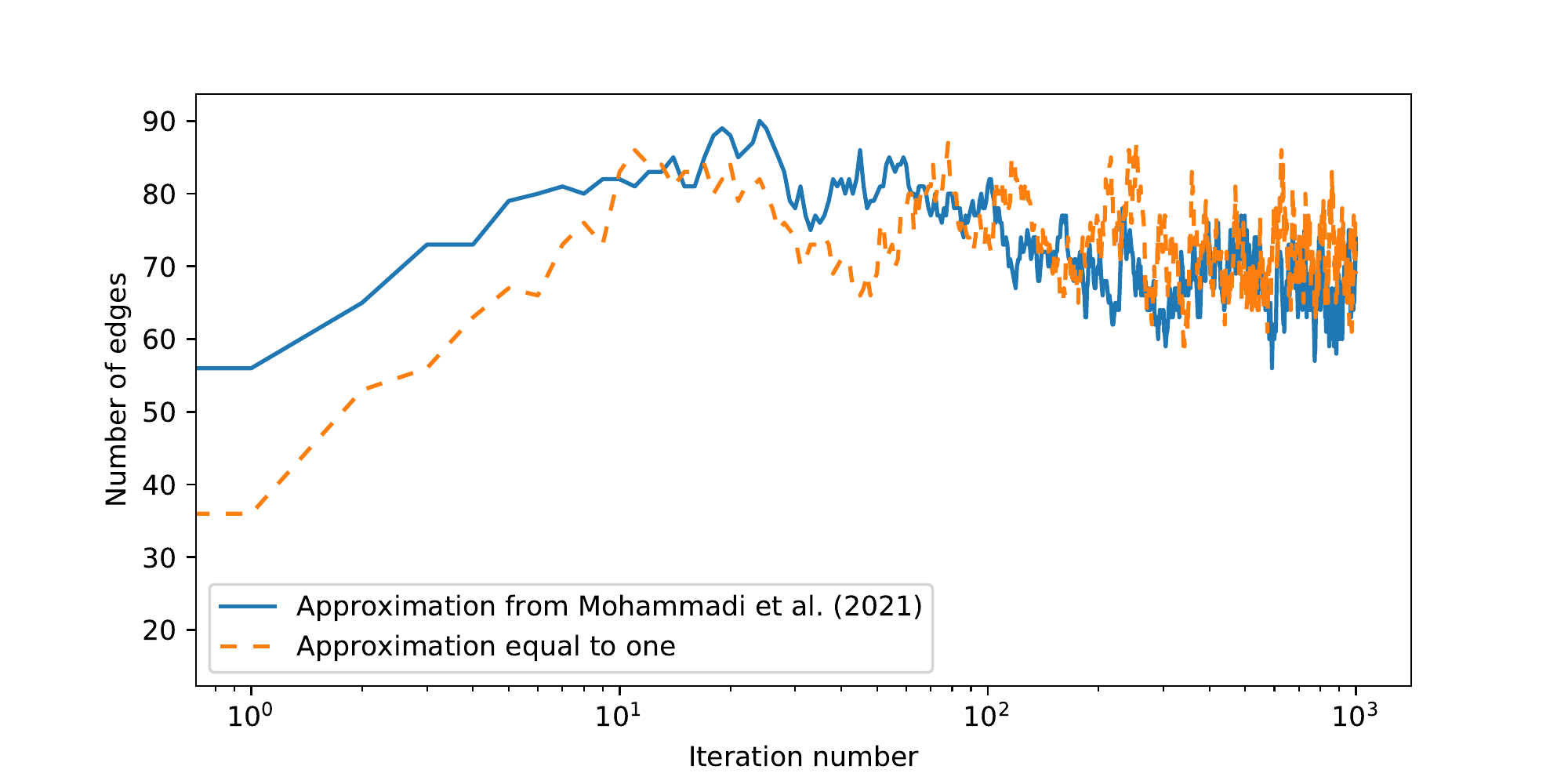}
\caption{
Trace plots for the number of edges from WWA
with the approximation
$\widehat{I_G/I_{\widetilde{G}}}$
as in \citet{Mohammadi2021}
and with
$\widehat{I_G/I_{\widetilde{G}}} = 1$
run on simulated data.
}
\label{fig:convergence}
\end{figure}

Figure~\ref{fig:convergence}
visualises the resulting MCMC chains.
The chain with the approximation from
\citet{Mohammadi2021}
converges away from the
initialisation with zero edges
slightly faster
than
$\widehat{I_G/I_{\widetilde{G}}} = 1$.

\section{Graph Decomposition and the \texorpdfstring{$G$}{G}-Wishart Sampler}
\label{sec:rgwish}

We investigate the effect of using graph decomposition on the computational cost of sampling from the $G$-Wishart distribution.
The two $G$-Wishart samplers considered are the one described in
Section~\ref{sec:wwa_complete} with graph decomposition
and the one proposed by \citet{Lenkoski2013}
without graph decomposition.
We record the computing time to sample from
$\mathcal{W}_G(\delta, D)$
for different $G$
with $\delta = 3$
and $D$ as in Section~2.3 of \citet{Dobra2011b}.
That is,
$D = I_p + 100A^{-1}$
with
$A$ given by $A_{ii}=1$ for $i=1,\dots,p$,
$A_{ij} = 0.5$ for $|i-j|=1$, $A_{1p}=A_{p1}=0.4$ and all other elements of $A$ being equal to zero.
The three types of graphs $G$ that we consider are
(i)
a decomposable graph
created by adding $p-3$ chords to a cycle,
specifically,
$E = \{(i, i+1)\mid i=1,\dots,p-1\}\cup \{(1,p)\}\cup\{(1,i)\mid i=3,\dots,p-1\}$;
(ii)
random graphs with independent edges
with edge inclusion probability $\rho = 0.5$
and
(iii) $\rho = 2/(p-1)$ such that the expected number of edges equals $p$.
We show results for 100 replicates and for $p=10,20,40,80$.

\begin{figure}[tbp]
\centering
\includegraphics[width=\textwidth]{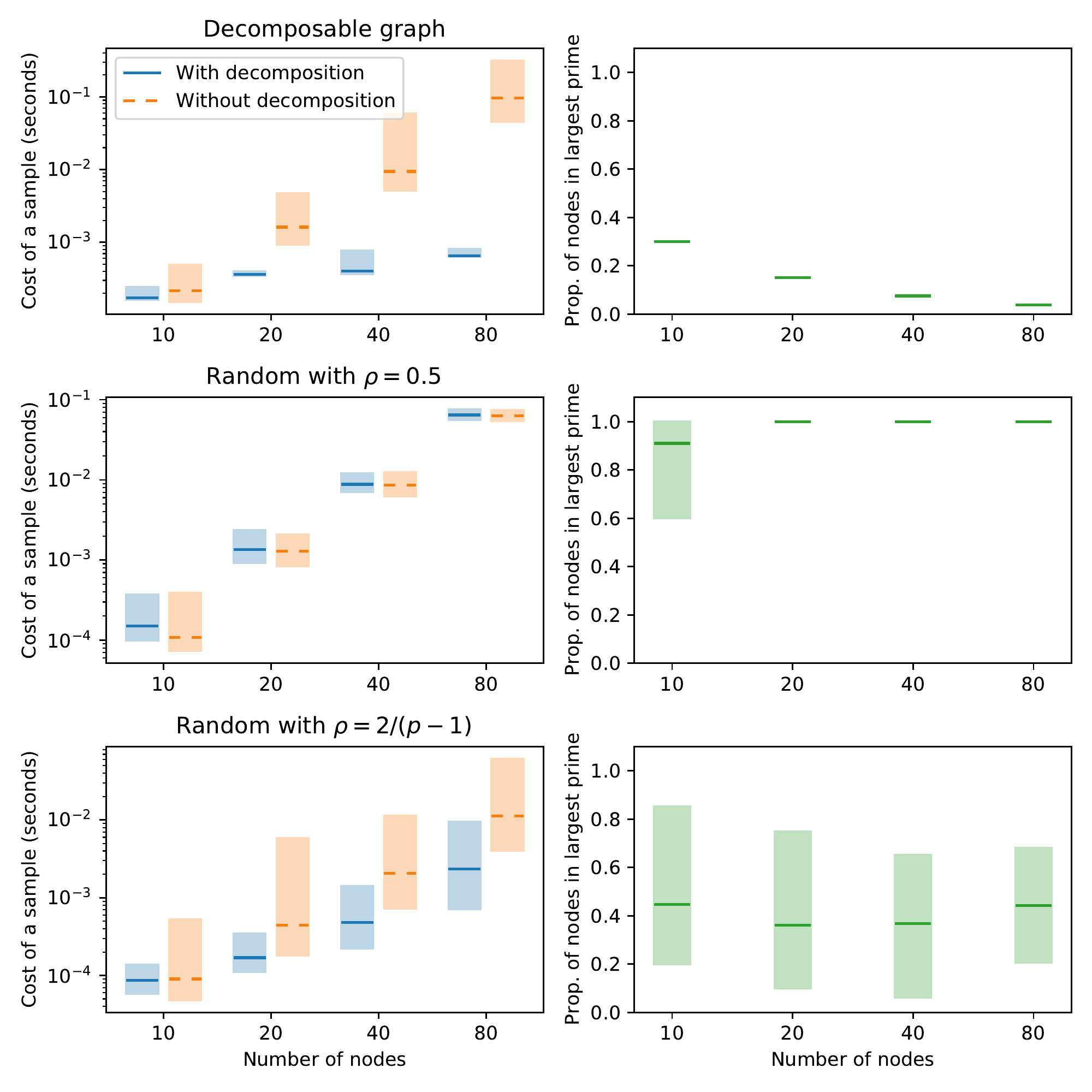}
\caption{ 
Cost of sampling from the $G$-Wishart distribution for the sampler with and without graph decomposition (left), and the proportion of nodes part of the graph's largest prime component (right) versus the number of nodes for the three types of graphs described in Section~\ref{sec:rgwish}.
The lines represent medians over 100 repetitions.
The boundaries of shaded areas correspond to the 2.5\% and 97.5\% quantiles.
}
\label{fig:rgwish}
\end{figure}

Figure~\ref{fig:rgwish}
shows that the $G$-Wishart sampler with graph decomposition
can take considerably less computing time
than the sampler without decomposition
if the graph's largest prime component has substantially
fewer nodes
than the graph itself. When no effective decomposition is available,
the additional cost incurred by attempting it
is negligible in these simulations.
Graphs sampled uniformly from the graph space $(\rho = 0.5)$
consist of a single prime component
with high probability
when the number of nodes is greater than or equal to $20$.
Thus, the use of graph decomposition does not yield a faster
$G$-Wishart sampler
for such graphs.
}

\section{Algorithm from \texorpdfstring{\citet{Cheng2012}}{Cheng and Lenkoski (2012)}}
\label{ap:cl}

We describe the CL algorithm of \citet{Cheng2012} in Algorithm~\ref{alg:cl} to aid the discussion in Section~\ref{sec:related}.
To highlight connections with WWA, we use the delayed acceptance framework to describe the algorithm even though \citet{Cheng2012} do not use such terminology.

The first accept-reject step in Step~\ref{step:delayed_accept_cl} of Algorithm~\ref{alg:cl} uses $\widehat{R}_\textnormal{CL}$ as odds instead of acceptance ratio.
This \add{corresponds to} Barker's algorithm \citep{Barker1965}:
write $\widehat{R}_\textnormal{CL} = \widetilde{z}/z$ for some positive numbers $z$ and $\widetilde{z}$ that correspond with the current and proposed state, respectively. Denote the acceptance probability by $\alpha$.
Accepting with odds $\widehat{R}_\textnormal{CL}$ means
$\widehat{R}_\textnormal{CL} = \alpha/(1-\alpha)$. Then, $\alpha = \widetilde{z} / (z + \widetilde{z})$ which is Barker's acceptance probability. Using $\widehat{R}_\textnormal{CL}$ as acceptance ratio would yield the Metropolis-Hastings acceptance probability $\alpha = 1\wedge \widehat{R}_\textnormal{CL}$.

\begin{algorithm}
\caption{\citep{Cheng2012} A Single MCMC Step of the CL Algorithm. \label{alg:cl}}
\textbf{Input:} Precision matrix $K$ and graph $G$.

\textbf{Output:} MCMC update for $(G, K)$
such that the invariant distribution is approximately the posterior $p(G, K\mid  Y)$.

\begin{enumerate}
    \item
    For each edge $e \in {\{(i,j)\mid 1\leq i<j\leq p \}}$, do the following:
    \begin{enumerate}
    	\item
    	Let $\widetilde{G}=(V,\widetilde{E})$ where $\widetilde{E} = E\cup\{e\}$ if $e\notin E$
    	and $\widetilde{E} = E\setminus\{e\}$ otherwise.
    	\item
    	Reorder the nodes in $G$ and $\widetilde{G}$ so that $e$ connects node\add{s}
    	$p-1$ and $p$. Rearrange $D$, $D^\star$ and $K$ accordingly. Denote the resulting quantities \add{by} a superscript $e$.
    	\item \label{step:delayed_accept_cl}
        Denote the upper triangular Cholesky decomposition of $K^e$ by $\Phi^e$.
    	`Promote' $\widetilde{G}$ to be considered for delayed acceptance with odds $\widehat{R}_\textnormal{CL}$ where
    	\[
    		\widehat{R}_\textnormal{CL} = \frac{p(\widetilde{G})}{p(G)}
    		N(\Phi^e_{-f}, D^{\star,e})^{|\widetilde{E}| - |E|}
    	\]
    	with
    	$N(\Phi^e_{-f}, D^{\star,e})$ given by Equation~\eqref{eq:N}.
    	If $\widetilde{G}$ is promoted:
    	\begin{enumerate}
        	\item
        	\label{step:dmh}
        	Generate a $\widetilde{K}^{0,e}$ by running the \add{maximum clique} block Gibbs sampler \citep{Wang2012} initialised at $K^e$ and with $\mathcal{W}_{\widetilde{G}^e}(\delta, D^e)$
        	as stationary distribution.
        	\item
        	Compute the upper triangular Cholesky decomposition $\widetilde{\Phi}^{0,e}$ of $\widetilde{K}^{0,e}$.
        	\item
        	Set $G=\widetilde{G}$ and $K = \widetilde{K}$ w.p.~$1\wedge N(\widetilde{\Phi}^{0,e}_{-f}, D^{e})^{|E| - |\widetilde{E}|}$.
    	\end{enumerate}
    	\item
    	Update $K$ by resampling $\Phi^e_{p-1,p}$ and $\Phi^e_{pp}$ according to $G^e$.
    \end{enumerate}
    \item
    Resample $K$ by running the \add{maximum clique} block Gibbs sampler \citep{Wang2012} with $\mathcal{W}_{G}(\delta^\star, D^\star)$ as stationary distribution.
\end{enumerate}
\end{algorithm}

\add{
\section{Quality of Posterior Approximation}
\label{sec:iris_and_cycle}

Section~\ref{sec:empirical} in the main text
focuses on MCMC efficiency.
Here, we additionally assess the performance of WWA
in terms of estimation of posterior summaries
such as posterior edge inclusion probabilities
and accuracy of precision matrix estimation.
While comparisons of effective sample size
might be hard to interpret across MCMC algorithms that have different target distributions,
such issues do not arise for posterior summaries.
Therefore,
we additionally compare with the CL algorithm detailed in Algorithm~\ref{alg:cl}
and the methodology of \citet{Mohammadi2019} as implemented in the R package \texttt{BDgraph}
which do not exactly have $p(G\mid Y)$ as invariant distribution.
For instance,
Step~\ref{step:dmh} of Algorithm~\ref{alg:cl}
does not sample from $\mathcal{W}_{\widetilde{G}^e}(\delta, D^e)$ directly,
resulting in an approximate version of the exchange algorithm, namely the double Metropolis-Hastings sampler
\citep{Liang2010},
which does not preserve the invariant distribution of the Markov chain.
As with DCBF in Section~\ref{sec:empirical},
we slightly modify the CL algorithm
by
executing the steps in Algorithm~\ref{alg:cl}
for $p$ edges drawn uniformly at random with replacement
instead of
all $m_{\max}$ possible edges
for a fairer comparison with WWA.

\subsection{Fisher's Iris Virginica Data}
\label{sec:iris}

We consider the example in \citet{Roverato2002}, \citet{AtayKayis2005} and \citet{Lenkoski2013}.
The data $Y$ consist of
$p=4$ demeaned measurements in centimetres of
sepal length, sepal width, petal length and petal width of
$n = 50$
Iris virginica plants.
The prior set-up is the same as in Section~\ref{sec:uniform}.
We run WWA,
DCBF,
WWA with delayed acceptance but without the informed proposal,
the CL algorithm and \texttt{BDgraph}.
The Markov chains are initialised at the empty graph
which is the default in \texttt{BDgraph}.
We run the chains for $10^3$ burn-in iterations followed by $10^6$ recorded iterations.

\begin{table}[tbp]
\caption{Posterior edge inclusion probabilities for the Iris virginica data as estimated by various MCMC methods.  \label{tab:iris}}
\begin{center}
\begin{tabular}{r|cccccc}
Edge & SL--SW & SL--PL & SL--PW & SW--PL & SW--PW & PL--PW \\\hline
WWA & 0.822 & 1.000 &  0.406 &   0.499 & 0.987 &  0.533 \\
DCBF & 0.822 & 1.000 & 0.405 &  0.501 & 0.987 &  0.533 \\
No informed proposal & 0.820 & 1.000 & 0.405 &  0.500 & 0.987 &  0.529 \\
CL algorithm & 0.820 & 1.000 & 0.402 & 0.498 & 0.987 & 0.529 \\
\texttt{BDgraph} & 1.000 & 1.000 & 1.000 & 1.000 & 1.000 & 1.000
\end{tabular}
\end{center}
Abbreviations: SL, sepal length; SW, sepal width; PL, petal length; PW, petal width
\end{table}

Table~\ref{tab:iris} presents the resulting estimates of the posterior edge inclusion probabilities.
Comparing these values with those in Table~1 of \citet{Lenkoski2013}
shows that all methods except for \texttt{BDgraph} provide accurate inclusion probabilities.
The approximate nature of the CL algorithm does not substantially affect these estimates,
though it seems to slightly underestimate the inclusion probabilities for the edges between
sepal length and petal width, and between sepal width and petal length.
The CL algorithm has the lowest cost of an independent sample, as defined in Section~\ref{sec:empirical}, at $0.31$ milliseconds
compared to $0.52$, $0.49$ and $0.72$ milliseconds
for WWA, DCBF and WWA without the informed proposal, respectively.
We do not compute cost of an independent sample for \texttt{BDgraph} as it performs continuous-time MCMC such that we cannot derive the integrated autocorrelation time in the same fashion as for the other algorithms.

\subsection{Cycle Graphs}
\label{sec:cycle_ap}

Here, we consider the same prior set-up and data generation as in Section~\ref{sec:cycle} for $p=10$ and $p=100$ nodes.
We apply the same MCMC set-ups as in Section~\ref{sec:iris} for $p=10$.
For $p=100$,
we use $10^4$ instead of $10^3$ burn-in iterations.
Because of their high computational cost per iteration,
we do not consider WWA with the informed proposal and DCBF,
and run WWA without the informed proposal
and \texttt{BDgraph}
for $10^5$ instead of $10^6$ recorded iterations
for $p=100$.

\begin{table}[tbp]
\caption{Performance measures with the set-up from Section~6.2 of \citet{Wang2012} for various MCMC methods.  \label{tab:cycle_ap}}
\begin{center}
\begin{tabular}{r|cccccc}
Measure & WWA & DCBF & No informed prop. & CL algorithm & \texttt{BDgraph} \\\hline\hline
 & \multicolumn{5}{c}{\shortstack{\vphantom{1cm}\\ $p = 10$ nodes}} \\\hline
Max.\ diff. & \textemdash & $6.0\cdot 10^{-3}$ & $7.7\cdot 10^{-3}$ & $6.7\cdot 10^{-3}$ & $6.3\cdot 10^{-2}$ \\
MSD & \textemdash & $3.5\cdot 10^{-6}$ & $6.4\cdot 10^{-6}$ & $6.1\cdot 10^{-6}$ & $3.4\cdot 10^{-4}$ \\
Min.\ inc.\ prob. & 0.133 & 0.134 & 0.132 & 0.137 & 0.174 \\
Max.\ inc.\ prob. & 0.997 & 0.997 & 0.997 & 0.997 & 0.994 \\
$p(G_\textnormal{true}\mid Y)$ & $2.1\cdot 10^{-5}$ & $8.0\cdot 10^{-6}$ & $3.6\cdot 10^{-5}$ & $2.3\cdot 10^{-5}$ & $2.8\cdot 10^{-5}$ \\
KL $\hat{K}$ & 8.2 & 8.2 & 8.2 & 8.2 & 8.4 \\
Frobenius $\hat{K}$ & 63 & 63 & 63 & 63 & 63 \\
CIS (millisec.) & 4.0 & 18 & 4.8 & 2.9 & \textemdash \\\hline
 & \multicolumn{5}{c}{\shortstack{\vphantom{1cm}\\ $p = 100$ nodes}} \\\hline
Max.\ diff. & & & \textemdash & $0.24$ & $0.24$ \\
MSD & & & \textemdash & $2.2\cdot 10^{-4}$ & $2.1\cdot 10^{-4}$ \\
Min.\ inc.\ prob. & & & 0.761 & 0.755 & 0.834 \\
Max.\ inc.\ prob. & & & 0.239 & 0.245 & 0.168 \\
$p(G_\textnormal{true}\mid Y)$ & & & $5.0\cdot 10^{-3}$ & $3.8\cdot 10^{-3}$ & $8.4\cdot 10^{-3}$ \\
KL $\hat{K}$ & & & 91 & 92 & 93 \\
Frobenius $\hat{K}$ & & & 2480 & 2480 & 2480 \\
CIS (seconds) & & & 33 & 33 & \textemdash \\\hline
\end{tabular}
\end{center}
\end{table}

Table~\ref{tab:cycle_ap} presents the results for the following performance measures:
\begin{description}
\item[Max.\ diff.] Maximum absolute difference from WWA of the posterior edge inclusion probabilities.
\item[MSD] Mean squared difference from WWA of the posterior edge inclusion probabilities.
\item[Min.\ inc.\ prob.] Minimum posterior inclusion probability for the edges included in the true underlying cycle graph.
\item[Max.\ inc.\ prob.] Maximum posterior inclusion probability for the edges excluded in the true underlying cycle graph.
\item[$\mathbf{p(G_\textnormal{true}\mid Y)}$] Posterior probability of the true underlying cycle graph.
\item[KL $\mathbf{\hat{K}}$] Kullback-Leibler divergence
$\textnormal{KL} = 0.5 \cdot \{ \mathrm{tr}(K_\textnormal{true}^{-1} \hat{K}) - p - \log(|\hat{K}| / |K_\textnormal{true}|) \}$
from the true data-generating distribution $\mathcal{N}(0_{p\times 1},\, K_\textnormal{true}^{-1})$ to $\mathcal{N}(0_{p\times 1},\, \hat{K}^{-1})$
where $\hat{K}$ is the posterior mean of the precision matrix \citep{Mohammadi2015}.
\item[Frobenius $\mathbf{\hat{K}}$] Frobenius norm of the matrix $\hat{K} - K_\textnormal{true}$.
\item[CIS] Cost of an independent sample in seconds. For the CL algorithm with $p=100$, CIS is computed from the last $10^5$ iterations for fair comparison.
\end{description}

The measures on the posterior inclusion probabilities do not show major differences between the MCMCs considered.
The `Min.\ inc.\ prob.', `Max.\ inc.\ prob.'\ and, for $p = 10$, `Max.\ diff.'\ of \texttt{BDgraph} differ from the other algorithms though not as extremely as the edge inclusion probabilities in Table~\ref{tab:iris}. This suggests that the approximations in \texttt{BDgraph} have an impact on the estimates of the posterior edge inclusion probabilities.

The median probability graph does not recover the true underlying cycle graph for $p=10$,
suggesting that $n=15$ observations do not provide enough information.
This contrasts with $p=100$ where $n=150$, which recovers the cycle graph with the median probability graph as based on `Min.\ inc.\ prob.' and `Max.\ inc.\ prob.' in line with the results reported in Section~6.2 of \citet{Wang2012} for the same simulation set-up.
The estimates of the posterior probabilities of the true graph $p(G_\textnormal{true}\mid Y)$ are small which is unsurprising given the uncertainty in the posterior and the size of the graph space.

The measures of accuracy of the posterior mean $\hat{K}$
for estimation of the precision matrix $K_\textnormal{true}$
do not vary notably across algorithms.
Note that Table~5 of \citet{Mohammadi2015}
reports substantially smaller `KL $\hat{K}$' for the posterior mean
of the Bayesian model that we consider
than for the graphical lasso estimate of the precision matrix.
Table~\ref{tab:cycle_ap} suggests that such superior performance for the precision matrix is also achieved when using approximate MCMC such as provided by \texttt{BDgraph}.

Finally,
we remark that the MCMC efficiency of WWA as measured by CIS is competitive with the CL algorithm.
That is,
WWA performs exact MCMC with a computational efficiency comparable to the CL algorithm
which does not  have the correct posterior as invariant distribution.
}

\bibliographystyle{chicago}
\bibliography{graph}